\documentclass[%
 reprint,
superscriptaddress,
 amsmath,amssymb,
 aps,
onecolumn,
showkeys,
]{revtex4-2}
\usepackage{graphicx}
\usepackage{dcolumn}
\usepackage{bm}

\begin{document}

\title{Predictions for Bottomonium from a Relativistic Screened Potential Model}

\author{Chaitanya Anil Bokade}
 \email{anshul497@gmail.com}
 \altaffiliation[ORCID iD: ]{0009-0007-5463-6812}
\author{Bhaghyesh}%
\email{bhaghyesh.mit@manipal.edu; Corresponding author}
 \altaffiliation[ORCID iD: ]{0000-0003-3994-9945}
\affiliation{Department of Physics, Manipal Institute of Technology\\Manipal Academy of Higher Education, Manipal, 576104, Karnataka, India}

\begin{abstract}
In this work, a comprehensive analysis of the mass spectra and decay properties of bottomonium states using a relativistic screened potential model is carried out. The mass spectrum, decay constants, $E1$ transitions, $M1$ transitions, and annihilation decay widths are evaluated. The interpretation of $\Upsilon(10355)$, $\Upsilon(10580)$,$\Upsilon(10860)$, and $\Upsilon(11020)$ as $S-D$ mixed bottomonium states are analysed. The $\Upsilon(10355)$ state is considered to be $3S-2D$, $\Upsilon(10580)$ and $\Upsilon(10753)$ are considered to be $4S-3D$ mixed states, and the $\Upsilon(10860)$ and $\Upsilon(11020)$ are deemed to be $5S-4D$ mixed states.


\end{abstract}
\maketitle


\section{\label{sec:Introduction} Introduction}

The study of heavy quarkonium, specifically bottomonium, has emerged as a captivating and influential field in contemporary particle physics. The allure of this research lies not only in the experimental endeavors aimed at unraveling the intricate properties of these heavy quark systems \cite{1,2}, but also its rich theoretical framework which allows to understand intricate interplay of perturbative and non-perturbative quantum chromodynamics (QCD) phenomena across a broad energy spectrum \cite{3}. The history of bottomonium traces back to $\Upsilon(1S)$, first discovered by the E288 Collaboration at Fermilab along with $\Upsilon(2S)$ and $\Upsilon(3S)$ \cite{4,5}. In 1982, the $\chi_{bJ}(2P)$ states were observed from CUSB detector at CESR in $\Upsilon(3S) \rightarrow \gamma \chi_{bJ}(2P)$  for $(J = 0,1,2)$ \cite{6,7}. The $\chi_{bJ}(1P)$ states were discovered in 1983 in $\eta_{b}(2S)\rightarrow \gamma \chi_{bJ}(1P)$ and $\chi_{bJ}(1P) \rightarrow \gamma \eta_{b}(1S)$ \cite{8}, and were later confirmed in the same year by CESR in $\Upsilon(2S) \rightarrow \gamma \chi_{bJ}(1P) \rightarrow \gamma \gamma \Upsilon(1S)$ \cite{9}. In 2005, most precise measurement of branching fractions and photon energies of $\chi_{bJ}(1P)$ and $\chi_{bJ}(2P)$ were conducted by the CLEO Collaboration \cite{10}. For the first time in 1980, CESR observed a peak above the $B\bar{B}$ threshold and suggested it as $\Upsilon(4S)$ \cite{11}. Later in 1985, CLEO at CESR, in addition to $\Upsilon(4S)$, reported the observation of $\Upsilon(10860)$ and $\Upsilon(11020)$ resonances \cite{12}. Most recently, Belle Collaboration in 2019 measured $e^{+}e^{-} \rightarrow \Upsilon(1,2,3 S) \pi^{+}\pi^{-}$ cross sections, determining masses and widths of $\Upsilon(10860)$ and $\Upsilon(11020)$ with improved precision \cite{13}. In 2004, CLEO Collaboration observed the $\Upsilon(1D)$ state at $10161.1 \pm 0.6 \pm 1.6$ MeV via a photon cascade of $\Upsilon(3S)$ decays, identifying it as $\Upsilon_{2}(1D)$ state \cite{14}. BABAR Collaboration later confirmed the $\Upsilon_{J}(1D)$ triplet in $\Upsilon(3S) \rightarrow \gamma \gamma \Upsilon(1D) \rightarrow \gamma \gamma \pi^{+}\pi^{-} \Upsilon(1S)$, with significance of $5.8\sigma$ for $\Upsilon_{2}(1D)$, while the significance for $\Upsilon_{1}(1D)$ and $\Upsilon_{3}(1D)$ states were very low \cite{15}. In 2008, BABAR discovered $\eta_{b}(1S)$ with $10\sigma$ significance via  $\Upsilon(3S) \rightarrow \gamma \eta_{b}(1S)$ \cite{16}. It was later confirmed by CLEO Collaboration \cite{17,18}, which also identified $\eta_{b}(2S)$ with $5\sigma$ significance in $\Upsilon(2S) \rightarrow \gamma \eta_{b}(2S)$ \cite{18}. Belle Collaboration observed $\eta_{b}(2S)$ for the first time in $h_{b}(2P) \rightarrow \gamma \eta_{b}(2S)$, measuring its mass as $9999.0 \pm 3.5_{-1.9}^{+2.8}$ MeV and hyperfine splitting as $m[\Upsilon(2S)]-m[\eta_{b}(2S)] = 24.3_{-4.5}^{+4.0}$ MeV \cite{19}. In 2011, BABAR Collaboration observed $h_{b}(1P)$ with $3.1\sigma$ significance in $\Upsilon(3S) \rightarrow \pi^{0} h_{b}(1P) \rightarrow \pi^{0} \gamma \eta_{b}(1S)$ \cite{20}. Belle Collaboration later confirmed $h_{b}(1P)$ in $\Upsilon(5S) \rightarrow \pi^{+}\pi^{-} h_{b}(1P)$ and discovered $h_{b}(2P)$ with $11.2\sigma$ significance, having mass of $10259.8 \pm 0.6_{-1.0}^{+1.4}$ MeV \cite{21}. In 2012, ATLAS Collaboration observed $\chi_{b}(3P)$ in $\chi_{b}(nP) \rightarrow \gamma \Upsilon(1S,2S)$ \cite{22}, later it was also confirmed by D0 Collaboration with $5.6\sigma$ significance at mass barycentre of $10551 \pm 14 \pm 17$ MeV \cite{23}. In 2018,  CMS Collaboration observed $\chi_{b1}(3P)$ and $\chi_{b2}(3P)$ in $\gamma \Upsilon(3S)$ decay mode, measuring their masses as $10513.42 \pm 0.41 \pm 0.18$ MeV and $10524.02 \pm 0.57 \pm 0.17$ MeV, respectively, and a mass difference of $10.60 \pm 0.64 \pm 0.17$ MeV \cite{24}. In 2012, from the data on $\Upsilon(5S)$ decays to $\Upsilon(nS) \pi^{+}\pi^{-} (n = 1,2,3)$ and $h_{b}(mP) \pi^{+}\pi^{-} (m = 1,2)$, Belle Collaboration observed two charged structures, $Z_{b}(10610)$ and $Z_{b}(10650)$ \cite{25}. Due to their charged property, they cannot be described in conventional quarkonium picture and require a four-quark configuration description such as hadronic molecules \cite{26,27}, tetraquarks \cite{28,29}, etc. In 2019, BELLE Collaboration discovered $\Upsilon(10753)$ with $5.2\sigma$ significance in $e^{+}e^{-} \rightarrow \Upsilon(nS) \pi^{+}\pi^{-}$ with mass of $10752.7 \pm 5.9_{-1.1}^{+0.7}$ MeV and width of $35.5_{-11.3-3.3}^{+17.6+3.9}$ MeV \cite{30}. It was also identified through cross-section calculations by BABAR and BELLE experiments \cite{31}. Even with significant progress in experimental domain there is still lack of key details, such as the total width and mass values of higher resonance, branching ratios for significant decay modes, etc. Unlike the rich charmonium-like $XYZ$ sector, there are only few unconventional  bottomonium-like states (e.g. $Z_{b}(10610)$, $Z_{b}(10650)$) that has been discovered. No experimental evidence exists for $X_{b}$, the bottomonium counterpart of $X(3872)$ \cite{32}. The pursuit of similar exotic states in the bottomonium system, such as the $X_{b}$ whose existence is predicted in multiple models \cite{33,34}, holds promise for understanding the nature of the internal structure of $X(3872)$. Exotic hadron studies have predominantly relied on $e^{+}e^{-}$ annihilation experiments, exemplified by BESIII, Belle, BaBar, and CLEO \cite{3}. Belle II at SuperKEKB aims to achieve for a peak luminosity of $8 \times 10^{35} \text{cm}^{2} \text{s}^{-1}$ by 2025, with operations extending to 2027 to collect over $50 \text{ab}^{-1}$ of data \cite{3}. Following the LHCb Upgrade I, Run-3 data will be crucial, while the PANDA experiment on antiproton-nucleon interactions and upcoming super $\tau$ - charm factories offer promising avenues for exploring novel states \cite{2,35}. In view of the potential for discovering new states in the bottomonium sector, we have developed a relativistic screened potential model that has proven effective for charmonium \cite{charm}. The proposed model provides a robust theoretical framework by considering relativistic effects and the screening of the potential. Our model can facilitate the identification and characterization of new and exotic states within the bottomonium spectrum. 

In this paper we have conducted a comprehensive study of bottomonium  in a relativistic screened potential model. In section \ref{sec:Methodology}, we discuss the theoretical model used to describe the bottomonium bound system and the numerical approach used to solve the relativistic Schrodinger equation. Decay constants and various decays are discussed in section \ref{sec:Decay properties}. In section \ref{sec:S-D Mixing}, $S-D$ mixing of bottomonium states is discussed. In section \ref{sec:Results and Discussion} a thorough investigation of our evaluation and interpretation of bottomonium states are conducted, along with comparison between the experimental results and other theoretical models. In section \ref{sec:Conclusion} we present our conclusion.

\section{\label{sec:Methodology} Methodology}

A relativistic potential model is developed to investigate various bottomonium properties. We utilize the spinless Salpeter equation, which is a relativistic extension of the Schrodinger equation \cite{36}
\begin{equation}
	\label{eq:1}
	H = \sqrt{-\nabla_{q}^{2} + m_{q}^{2}}+\sqrt{-\nabla_{\bar{q}}^{2} + m_{\bar{q}}^{2}} + V(r) \,,
\end{equation}
where $\vec{r}= \vec{x}_{\bar{q}}-\vec{x}_{q}$, $\vec{x}_{\bar{q}}$ and $\vec{x}_{q}$  are the coordinates of the quarks and operators $\nabla_{q}^{2}$ and  $\nabla_{\bar{q}}^{2}$ are the partial derivatives of those coordinates respectively. $m_{q}$  and $m_{\bar{q}}$ are the masses of quark and an anti-quark respectively. The interaction potential $V(r)$ between the quark and antiquark is composed of two components: $V_{V}(r)$, representing the one-gluon-exchange coulomb potential term that is dominant at short distance, and $V_{S}(r)$, which represents the linear confining term adjusted to account for colour screening effects at longer distances \cite{37}:
\begin{align}
	V_{V}(r) &= -\frac{4}{3}\frac{\alpha_{s}(r)}{r} \,,
	\\
	V_{S}(r) &= \lambda\left(\frac{1 - e^{-\mu r}}{\mu}\right) + V_{0} \,,\\
	V(r) &= V_{V}(r)+V_{S}(r) \,.
\end{align}
Here $\lambda$ is the linear potential slope and $\mu$ is the screening factor which regulates the behaviour of the long-range component of $V(r)$, causing it to flatten out as $r$ becomes much larger than $1/\mu$ and exhibit a linear increase as $r$ becomes much smaller than $1/\mu$. The $V(r)$ converges to Cornell potential as $\mu \rightarrow 0$ \cite{37}. $\alpha_{s}(r)$ is the running coupling constant in coordinate space obtained by Fourier transformation of coupling constant in momentum space $\alpha_{s}\left(Q^{2}\right)$ \cite{36} and is given by
\begin{equation}
	\alpha_{s}(r) = \sum_{i}\alpha_{i}\frac{2}{\sqrt{\pi}}\int_{0}^{\gamma_{i}r}e^{-x^{2}}dx \,,
\end{equation}
where $\alpha_{i}'s$ are the free parameters to imitate the short-distance behaviour of $\alpha_{s}\left(Q^{2}\right)$ as predicted by QCD. The parameters values are taken as $\alpha_{1}=0.15$, $\alpha_{2}=0.15$, $\alpha_{3}=0.20$, and $\gamma_{1}=1/2$, $\gamma_{2}=\sqrt{10}/2$, $\gamma_{3}=\sqrt{1000}/2$ \cite{38}. The Hamiltonian $H$ is solved as an eigenvalue equation using the method developed in \cite{38,39}. The Hamiltonian Eq. \eqref{eq:1} as a eigen equation is given by
\begin{equation}
	E\Psi(\vec{r}) = 
	\left[\sqrt{-\nabla_{q}^{2} + m_{q}}+\sqrt{-\nabla_{\bar{q}}^{2} + m_{\bar{q}}} + V(r)\right]\Psi(\vec{r}) \,.
\end{equation}
The wave function can be expanded using spectral integration, which allows us to express the wave function as an integral over the eigenstates of the Hamiltonian $H$:
\begin{equation}
	\Psi(\vec{r}) = \int d^{3}r'\int \frac{d^{3}k}{(2\pi)^{3}}e^{i\vec{k}(\vec{r}-\vec{r}')}\Psi(\vec{r}') \,.
\end{equation}
Eq. \eqref{eq:1} can be rewritten as
\begin{equation}
	\label{eq:2}
	E\Psi\left(\vec{r}\right) =
	\int d^{3}r'\frac{d^{3}k}{(2\pi)^{3}}\left(\sqrt{k^{2} + m_{q}}+\sqrt{k^{2} + m_{\bar{q}}}\right) 
	e^{i\vec{k}(\vec{r}-\vec{r}')}\Psi\left(\vec{r}'\right) + V(r)\Psi\left(\vec{r}\right) \,.
\end{equation}
The exponential term can be expanded in terms of spherical harmonics as
\begin{equation}
	\label{eq:2.5}
	e^{i\vec{k}\cdot\vec{r}} = 4\pi\sum_{nl}Y_{nl}^{*}\left(\hat{k}\right)Y_{nl}(\hat{r})j_{l}(kr)i^{l}\,,
\end{equation}
where $j_{l}$ is the spherical Bessel function, $Y_{nl}^{*}(\hat{k})$ and $Y_{nl}(\hat{r})$ are the spherical harmonics with normalization condition $\int d\Omega Y_{n_{1}l_{1}}(\hat{k})Y_{n_{2}l_{2}}(\hat{r}) = \delta_{n_{1}n_{2}}\delta_{l_{1}l_{2}}$, $\hat{k}$ and $\hat{r}$ are unit vectors along the $\vec{k}$ and $\vec{r}$ direction, respectively. The wave function can be factorized into radial $R_{l}(r)$ and angular $Y_{nl}(r)$ parts, and substituting Eq. \eqref{eq:2.5} in \eqref{eq:2} and simplifying, we get \cite{38,39} 
\begin{equation}
	\label{eq:3}
	Eu_{l}(r) = 
	\frac{2}{\pi}\int dkk^{2}\int dr'rr'\left(\sqrt{k^{2} + m_{q}^{2}}+\sqrt{k^{2} + m_{\bar{q}}^{2}}\right)
	j_{l}(kr)j_{l}(kr')u_{l}(r') + V(r)u_{l}(r) \,,
\end{equation}
where $u_{l}(r)$ is the reduced radial wave function
$(R_{l}(r) =  u_{l}(r)/r)$. 
When the separation distance grows for a quark-antiquark bound state, the wavefunction gradually decreases and eventually approaches zero at sufficiently large distance. In order to represent this behaviour, a characteristic distance scale $L$ is introduced, confining the bound state's wavefunction within the spatial interval of $0<r<L$. Next, one can expand the reduced wavefunction $u_l(r)$ in terms of spherical Bessel function for angular momentum $l$ as
\begin{equation}
	\label{eq:4}
	u_l(r) = \sum_{n=1}^{\infty}c_{n}\frac{a_{n}r}{L}j_{l}\left(\frac{a_{n}r}{L}\right) \,,
\end{equation}
where $c_{n}$'s are the expansion coefficients, $a_{n}$ are the $n$-th root of the spherical Bessel function, $j_{l}(a_{n})=0$. For large value of $N$  Eq. \eqref{eq:4} can be truncated. The momentum is discretized as a result of confinement of space, which allows us to replace $a_{n}/ L\leftrightarrow k$  and the integration in Eq. \eqref{eq:3} can be replaced by $\int dk \rightarrow \sum_{n}\Delta a_{n}/ L$, where,  $\Delta a_{n}= a_{n} - a_{n-1}$. For finite space interval, $0<r,r'<L$, incorporating all the changes in the Eq. \eqref{eq:3}, we get the final equation in terms of the coefficients $c_{n}$'s as \cite{38,39}
\begin{equation}
	\label{eq:5}
	Ec_{m} =
	\sum_{n=1}^{N}\frac{a_{n}}{N_{m}^{2}a_{m}}\int_{0}^{L}drV(r)r^{2}j_{l}\left(\frac{a_{m}r}{L}\right)\left(\frac{a_{n}r}{L}\right)c_{n}
	+\frac{2}{\pi L^{3}}\left[\sqrt{\left(\frac{a_{m}}{L}\right)^{2} + m_{q}^{2}}+\sqrt{\left(\frac{a_{m}}{L}\right)^{2} + m_{\bar{q}}^{2}}\right]
	\Delta a_{m}a_{m}^{2}N_{m}^{2}c_{m} \,,
\end{equation}
where $N_{m}$ is module of spherical Bessel function
\begin{equation}
	N_{m}^{2} = \int_{0}^{L}dr'r'^{2}j_{l}\left(\frac{a_{m}r'}{L}\right)^{2} \,.
\end{equation}
When $L$ and $N$ attain sufficiently large values, the solution tends to become nearly stationary \cite{38,39}. The spin dependent interaction potential is given by \cite{40,41}
\begin{equation}
	V_{SD}(r) = V_{SS}(r)\vec{S}_{q}\cdot \vec{S}_{\bar{q}}+V_{LS}(r)\vec{L}\cdot \vec{S}+V_{T}(r)S_{12} \,.
\end{equation}
$V_{SS}$ is the spin singlet-triplet hyperfine splitting term given by
\begin{equation}
	V_{SS}(r) = \frac{32\pi\alpha_{s}(r)}{9m_{q}^{2}}\tilde{\delta}_{\sigma}(r) \,.
\end{equation}
Here $\tilde{\delta}_{\sigma}(r) = \left(\frac{\sigma}{\sqrt{\pi}}\right)^{3}e^{-\sigma^{2}r^{2}}$ is the smeared delta function \cite{42,43}. To regularize the non-zero hyperfine splitting, smearing of delta function as Gaussian of width $1/\sigma$ is necessary \cite{42,43}. The spin orbit term $V_{LS}$ and the tensor term $V_{T}$ which describe the fine structure splitting of the states are given by
\begin{align}
	V_{LS}(r) &= \frac{1}{2m_{q}^{2}r}\left(3V_{V}^{'}(r)-V_{S}^{'}(r)\right) \,,
	\nonumber \\
	V_{T}(r) &= \frac{1}{m_{q}^{2}}\left(\frac{V_{V}^{'}(r)}{r}-V_{V}^{''}(r)\right) \,. 
\end{align}
The tensor operator $S_{12}=3\left(\vec{S}_{q}\cdot\hat{r} \right)\left(\vec{S}_{\bar{q}}\cdot\hat{r}\right)- \vec{S}_{q}\cdot\vec{S}_{\bar{q}}$, has non-vanishing diagonal matrix elements only between $L > 0$ spin-triplet states. The spin-dependent interactions are diagonal in a $|J,L,S\rangle$ basis with matrix elements given by \cite{42,43,44}
\begin{align}
	\langle\vec{S}_{q}\cdot \vec{S}_{\bar{q}}\rangle &= \frac{1}{2}S^{2}-\frac{3}{4} \,,
	\nonumber \\
	\langle\vec{L}\cdot \vec{S}\rangle &= \frac{1}{2}\left[J(J+1)-L(L+1)-S(S+1)\right] \,,
	\nonumber \\
	\langle S_{12} \rangle &= \begin{cases}
		-\frac{L}{6(2L+3)} & J=L+1 
		\\
		\frac{1}{6} & J=L
		\\
		-\frac{L+1}{6(2L-1)} & J=L-1 \,.
	\end{cases} 
\end{align}

\noindent Eq. \eqref{eq:5} represents an eigenvalue equation in matrix form, which is solved numerically. The eigenvalues correspond to the masses of spin-averaged states and the eigenvectors represent their wave functions. Using the obtained normalized wave functions for the spin-averaged states, the spin-dependent corrections are evaluated perturbatively.  
The model parameters are determined using the $\chi^{2}$ fit method  through minimizing the $\chi^{2}$, defined as
\begin{equation}
	\chi^{2} = \sum_{i}\left(\frac{M^{i}_{Exp}-M^{i}_{Th}}{M^{i}_{Er}}\right)^{2}\,,
\end{equation}
where $M^{i}_{Exp}$ and $M^{i}_{Th}$ are the experimental mass and predicted mass respectively, and $M^{i}_{Er}$ is the error in $M^{i}_{Exp}$. The errors of observed masses $M^{i}_{Er}$ are taken as $0.1\%$ of the masses of the respective states, $M^{i}_{Exp}$. These errors are different from their corresponding experimental uncertainties,  which are too small for some states and are unevenly distributed. This approach ensures balanced weighting in the fitting process and prevents states which have lower experimental errors from disproportionately influencing the fit \cite{57}. For fitting we have considered the well established four $S$-wave states $\eta_b(1S,2S), \Upsilon(1S,2S)$, four $P$-wave states $h_c(1P,2P), \chi_{b1}(1P,2P)$ and one $D$-wave state, $1^{3}D_{2}$. Using this approach, we obtain a $\chi^{2}$ value of $14.1$. The fitted parameters are listed in Table \ref{tab:1}. The masses of $S,P,D,F$ and $G$ states are presented in Tables \ref{tab:2}-\ref{tab:5}, respectively.

\begin{table}
	\caption{\label{tab:1} Parameters used in our model}
	\begin{ruledtabular}
		\begin{tabular}{ccccc}
			$m_{q}$ (GeV) & $\sigma$ ($\text{GeV}^{2}$) & $\lambda$ (GeV) & $\mu$ (GeV) & $\Lambda$ (GeV)
			\\
			\colrule
			4.744 & 4.967 & 0.240 & 0.039 & 0.17
			\\
		\end{tabular}
	\end{ruledtabular}
\end{table}

\section{\label{sec:Decay properties} Decay properties}

\noindent Bottomonium decays are important for understanding internal structures, revealing underlying dynamics, and distinguishing states. Comparison of mass spectra and decay properties with experimental data helps to validate theoretical models. Using the obtained wave functions, we calculate various decay properites of bottomonium. 

Decay constants are fundamental parameters that characterize the strength of the weak interaction responsible for the decay processes, and it measures the probability amplitude to decay into lighter hadrons. The decay constant of pseudoscalar $(f_{P})$ and vector $(f_{V})$ states can be calculated using the Van Royen Weisskopf formula \cite{45}
\begin{equation}
	f_{P/V} = \sqrt{\frac{3|R_{P/V}(0)|^{2}}{\pi M_{P/V}}}\bar{C}(\alpha_{s}) \,,
\end{equation}
where $R_{P/V}(0)$ is the radial wavefunction at the origin for pseudoscalar (vector) meson state, $M_{P/V}$ is the mass of the pseudoscalar (vector) meson state and $\bar{C}(\alpha_{s})$ is the QCD correction given by \cite{46}
\begin{equation}
	\bar{C}^{2}(\alpha_{s}) = 1 - \frac{\alpha_{s}(\mu)}{\pi}\left(\delta^{P,V} - \frac{m_{q} - m_{\bar{q}}}{m_{q} + m_{\bar{q}}}\ln{\frac{m_{q}}{m_{\bar{q}}}}\right) \,,
\end{equation}
where $\delta^{P}=2$ and $\delta^{V}=8/3$. The decay constant of $P$-wave states can be evaluated using \cite{47,48} 
\begin{align}
	f_{\chi_{0}} = 12\sqrt{\frac{3}{8\pi m_{q} }}\left(\frac{|R'_{\chi_{0}}(0)|}{M_{\chi_{0}}}\right) \,,
	\nonumber \\
	f_{\chi_{0}} = 8\sqrt{\frac{9}{8\pi m_{q} }}\left(\frac{|R'_{\chi_{1}}(0)|}{M_{\chi_{1}}}\right) \,.
\end{align}
Here $M_{\chi_{0}}$ and $M_{\chi_{1}}$ are the masses of $\chi_{0}$ and $\chi_{1}$ states, respectively. The decay constants for pseudoscalar $f_{P}$ and vector $f_{V}$ are presented in Table \ref{tab:6} and decay constants for $f_{\chi_{0}}$ and $f_{\chi_{1}}$ are presented in Table \ref{tab:7}. Bottomonium annihilation decays leave distinct signals in experimental data, allowing bottomonium states to be identified and characterized in high-energy collider experiments and precision spectroscopic investigations. 

The leptonic decay formula for $S$-wave $(n^{3}S_{1})$  and $D$-wave $(n^{3}D_{1})$ states are calculated using Van Royen-Weisskopf formula along with the QCD correction factor \cite{45,49,50,51}
\begin{align}
	\Gamma\left(n^{3}S_{1} \rightarrow l^{+}l^{-}\right) &= \frac{4\alpha^{2}e_{q}^{2}}{M(n^{3}S_{1})^{2}}|R_{nS}(0)|^{2}
	\left[1-\frac{16\alpha_{s}(\mu)}{3\pi}\right] \,,
	\nonumber \\
	\Gamma\left(n^{3}D_{1} \rightarrow l^{+}l^{-}\right) &= \frac{25\alpha^{2}e_{q}^{2}}{2m_{q}^{4}M(n^{3}D_{1})^{2}}|R''_{nD}(0)|^{2} \,, 
\end{align}
where $R'_{nL}(0)$ is the value of radial wavefunction at origin for $nL$ state and $(')$ represents the order of derivative and $M(n^{2S+1}L_{J})$ is the mass of $n^{2S+1}L_{J}$ state. 

The annihilation decays for $S$-wave $(n^{1}S_{0})$  and $P$-wave $(n^{3}P_{0}\ \text{and}\ n^{3}P_{2})$ into two photons $(\gamma\gamma)$ and $S$-wave $(n^{3}S_{1})$ states into three photons $(\gamma\gamma\gamma)$ with first order QCD correction factors are given by \cite{49,50}
\begin{align}
	\Gamma\left(n^{1}S_{0} \rightarrow \gamma\gamma\right) &= \frac{2^{2}3\alpha^{2}e_{q}^{4}}{M(n^{1}S_{0})^{2}}|R_{nS}(0)|^{2}
	\left[1-\frac{3.4\alpha_{s}(\mu)}{\pi}\right] \,,
	\nonumber \\
	\Gamma\left(n^{3}P_{0} \rightarrow \gamma\gamma\right) &= \frac{2^{4}27\alpha^{2}e_{q}^{4}}{M(n^{3}P_{0})^{4}}|R'_{nP}(0)|^{2}
	\left[1+\frac{0.2\alpha_{s}(\mu)}{\pi}\right] \,, 
	\nonumber \\
	\Gamma\left(n^{3}P_{2} \rightarrow \gamma\gamma\right) &= \frac{2^{4}36\alpha^{2}e_{q}^{4}}{5M(n^{3}P_{2})^{4}}|R'_{nP}(0)|^{2}
	\left[1-\frac{16\alpha_{s}(\mu)}{3\pi}\right] \,,
	\nonumber \\
	\Gamma\left(n^{3}S_{1} \rightarrow \gamma\gamma\gamma\right) &= \frac{2^{2}4(\pi^{2}-9)\alpha^{3}e_{q}^{6}}{3\pi M(n^{3}S_{1})^{2}}|R_{nS}(0)|^{2}
	\left[1-\frac{12.6\alpha_{s}(\mu)}{\pi}\right] \,.
\end{align}

\noindent The annihilation decays for $S$-wave $(n^{1}S_{0})$, $P$-wave $(n^{3}P_{0}\ \text{and}\ n^{3}P_{2})$, $D$-wave $(n^{1}D_{2})$, $F$-wave $(n^{3}F_{2} ,n^{3}F_{3}\ \text{and}\ n^{3}F_{4})$ and $G$-wave $(n^{1}G_{4})$ states into two gluons $(gg)$ with first order QCD correction factors are given by \cite{49,50,52}
\begin{align}
	\Gamma\left(n^{1}S_{0} \rightarrow gg\right) &= \frac{2^{2}2\alpha_{s}^{2}(\mu)}{3M(n^{1}S_{0})^{2}}|R_{nS}(0)|^{2}
	\left[1+\frac{4.8\alpha_{s}(\mu)}{\pi}\right] \,, 
	\nonumber \\
	\Gamma\left(n^{3}P_{0} \rightarrow gg\right) &= \frac{2^{4}6\alpha_{s}^{2}(\mu)}{M(n^{3}P_{0})^{4}}|R'_{nP}(0)|^{2}
	\left[1+\frac{10\alpha_{s}(\mu)}{\pi}\right] \,,
	\nonumber \\
	\Gamma\left(n^{3}P_{2} \rightarrow gg\right) &= \frac{2^{4}8\alpha_{s}^{2}(\mu)}{5M(n^{3}P_{2})^{4}}|R'_{nP}(0)|^{2}
	\left[1-\frac{0.1\alpha_{s}(\mu)}{\pi}\right] \,,
	\nonumber \\
	\Gamma\left(n^{1}D_{2} \rightarrow gg\right) &= \frac{2^{6}2\alpha_{s}^{2}(\mu)}{3\pi M(n^{1}D_{2})^{6}}|R''_{nD}(0)|^{2}
	\left[1-\frac{2.2\alpha_{s}(\mu)}{\pi}\right] \,,
	\nonumber \\
	\Gamma\left(n^{3}F_{2} \rightarrow gg\right) &= \frac{2^{8}919\alpha_{s}^{2}(\mu)}{135 M(n^{3}F_{2})^{8}}|R'''_{nF}(0)|^{2} \,,
	\nonumber \\
	\Gamma\left(n^{3}F_{3} \rightarrow gg\right) &= \frac{2^{8}20\alpha_{s}^{2}(\mu)}{27 M(n^{3}F_{3})^{8}}|R'''_{nF}(0)|^{2} \,,
	\nonumber \\
	\Gamma\left(n^{3}F_{4} \rightarrow gg\right) &= \frac{2^{8}20\alpha_{s}^{2}(\mu)}{27 M(n^{3}F_{4})^{8}}|R'''_{nF}(0)|^{2} \,,
	\nonumber \\
	\Gamma\left(n^{1}G_{4} \rightarrow gg\right) &= \frac{2^{10}2\alpha_{s}^{2}(\mu)}{3\pi M(n^{1}G_{4})^{10}}|R''''_{nG}(0)|^{2} \,.
\end{align}

\noindent The annihilation decays for $S$-wave $(n^{3}S_{1})$, $P$-wave $(n^{1}P_{1})$ and $D$-wave $(n^{3}D_{1}, n^{3}D_{2}$ and $n^{3}D_{3})$ states into three gluons $(ggg)$ with first order QCD correction factors are given by \cite{49,50,53}
\begin{align}
	\Gamma\left(n^{3}S_{1} \rightarrow ggg\right) &=\frac{2^{2}10(\pi^{2}-9)\alpha_{s}^{3}(\mu)}{81\pi M(n^{3}S_{1})^{2}}|R_{nS}(0)|^{2}
	\left[1-\frac{4.9\alpha_{s}(\mu)}{\pi}\right] \,,
	\nonumber \\
	\Gamma\left(n^{1}P_{1} \rightarrow ggg\right) &= \frac{2^{4}20\alpha_{s}^{3}(\mu)}{9\pi M(n^{1}P_{1})^{4}}|R'_{nP}(0)|^{2}
	\ln\left(m_{q}\langle r\rangle \right) \,,
	\nonumber \\
	\Gamma\left(n^{3}D_{1} \rightarrow ggg\right) &= \frac{2^{6}760\alpha_{s}^{3}(\mu)}{81\pi M(n^{3}D_{1})^{6}}|R''_{nD}(0)|^{2}
	\ln\left(4m_{q}\langle r\rangle \right) \,,
	\nonumber \\
	\Gamma\left(n^{3}D_{2} \rightarrow ggg\right) &= \frac{2^{6}10\alpha_{s}^{3}(\mu)}{9\pi M(n^{3}D_{2})^{6}}|R''_{nD}(0)|^{2}
	\ln\left(4m_{q}\langle r\rangle \right) \,,
	\nonumber \\
	\Gamma\left(n^{3}D_{3} \rightarrow ggg\right) &= \frac{2^{6}40\alpha_{s}^{3}(\mu)}{9\pi M(n^{3}D_{3})^{6}}|R''_{nD}(0)|^{2}
	\ln\left(4m_{q}\langle r\rangle \right) \,.
\end{align}

\noindent The annihilation decays for $S$-wave $(n^{3}S_{1})$ states via strong and electromagnetic interaction, into a photon and two gluons $(\gamma gg)$ \cite{43,49,54} and the $P$ - wave $(n^{3}P_{1})$ into a light flavour meson and a gluon $(q\bar{q}g)$ are given by \cite{49,50}
\begin{align}
	\Gamma\left(n^{3}S_{1} \rightarrow \gamma gg\right) &=\frac{2^{2}8(\pi^{2}-9)e_{q}^{2}\alpha\alpha_{s}^{3}(\mu)}{9\pi M(n^{3}S_{1})^{2}}|R_{nS}(0)|^{2}
	\left[1-\frac{7.4\alpha_{s}(\mu)}{\pi}\right] \,,
	\nonumber \\
	\Gamma\left(n^{3}P_{1} \rightarrow q\bar{q}g\right) &= \frac{2^{4}8n_{f}\alpha_{s}^{3}(\mu)}{9\pi M(n^{3}P_{1})^{4}}|R'_{nP}(0)|^{2}\ln\left(m_{q}\langle r\rangle \right) \,,
\end{align}

\noindent where $n_{f}$ is the number of flavors. For all decays the strong coupling constant $\alpha_{s}(\mu)$ is calculated using the expression
\begin{equation}
	\alpha_{s}(\mu) = \frac{4\pi}{\beta_{0}\ln\frac{\mu^{2}}{\Lambda^{2}}}\left[1-\frac{\beta_{1}\ln\left(\ln\frac{\mu^{2}}{\Lambda^{2}}\right)}{\beta_{0}^{2}\ln\frac{\mu^{2}}{\Lambda^{2}}}\right] \,,
\end{equation}
where $\beta_{0}=11-(2/3)n_{f}$, $\beta_{1}=102- (18/3)n_{f}$, $\Lambda$ is the QCD constant taken from \cite{50}, and $\mu$ is the reduced mass. All annihilation decay widths for $b\bar{b}$ bound system are presented in Tables \ref{tab:8}-\ref{tab:14}, respectively. 

The bottomonium states have substantially heavier masses and their intrinsic compactness is more pronounced \cite{55}. As a result, radiative transitions in bottomonium are expected to be dominant because of the favourable conditions for photon emission or absorption \cite{55}. Radiative transitions serves as an effective means to detect, especially for states with higher quantum numbers that are hard to observe with traditional techniques. The $E1$ radiative partial widths between the states $\left(n_{i}^{2S+1}L_{J_{i}}^{i} \rightarrow \gamma+n_{f}^{2S+1}L_{J_{f}}^{f}\right)$ are given by \cite{56,57}
\begin{equation}
	\label{eq:6}
	\Gamma_{E1}\left(i \rightarrow \gamma+f\right) = \frac{4\alpha e_{q}^{2}}{3}E_{\gamma}^{3}
	\frac{E_{f}}{M_{i}}C_{fi}|\epsilon_{fi}|^{2}\delta_{S_{f}S_{i}} \,.
\end{equation}
Here $\alpha=1/137$ is the fine structure constant,  $e_{q}$ is the quark charge, $E_{f}$ is the energy of the final state, $M_{i}$ is the mass of the initial state, $E_{\gamma}= \left(M_{i}^{2}-M_{f}^{2}\right)/2M_{i}$ is the emitted photon energy. $M_{f}$ is the mass of the final state.  $E_{f}/M_{i}$  is the relativistic phase factor and $C_{fi}$ is the statistical factor given by 
\begin{equation}
	C_{fi} = \max\left(L_{i},L_{f}\right)\left(2J_{f}+1\right)
	\left\{
	\begin{array}{ccc}
		J_{i} & 1 & J_{f} \\
		L_{f} & S & L_{i} \\
	\end{array}
	\right\}^{2} \,,
\end{equation}
where $\{:::\}$ are the $6j$ symbol. In Eq. \eqref{eq:6}, $\epsilon_{fi}$ is the overlapping integral determined using the initial $R_{n_{i}l_{i}}(r)$ and final state $R_{n_{f}l_{f}}(r)$ wavefunctions:
\begin{equation}
	\epsilon_{fi} = \frac{3}{E_{\gamma}}\int_{0}^{\infty}drR_{n_{i}l_{i}}(r)R_{n_{f}l_{f}}(r)
	\left[\frac{E_{\gamma}r}{2}j_{0}\left(\frac{E_{\gamma}r}{2}\right)-j_{1}\left(\frac{E_{\gamma}r}{2}\right)\right] \,
\end{equation}
The $M1$ radiative partial widths between the states $\left(n_{i}^{2S_{i}+1}L_{J_{i}} \rightarrow \gamma+n_{f}^{2S_{f}+1}L_{J_{f}}\right)$ are given by \cite{42,56,58}
\begin{equation}
	\Gamma_{M1}\left(i \rightarrow \gamma+f\right) = \frac{4\alpha \mu_{q}^{2}}{3}\frac{2J_{f}+1}{2L+1}E_{\gamma}^{3}
	\frac{E_{f}}{M_{i}}
	|m_{fi}|^{2}\delta_{L_{f}L_{i}}\delta_{S_{f}S_{i}\pm1} \,,
\end{equation}
where $m_{fi}$ is given by
\begin{equation}
	m_{fi} = \int_{0}^{\infty}drR_{n_{i}l_{i}}(r)R_{n_{f}l_{f}}(r)\left[j_{0}\left(\frac{E_{\gamma}r}{2}\right)\right] \,,
\end{equation}
and $\mu_{q}$ is the magnetic dipole moment given by \cite{54}
\begin{equation}
	\mu_{q} = \frac{m_{\bar{q}}e_{q}-m_{q}e_{\bar{q}}}{2m_{q}m_{\bar{q}}} \,.
\end{equation}
The $E1$ transitions widths for $S, P, D, F\ \text{and}\ G$ wave states are presented in Tables \ref{tab:15}-\ref{tab:19}, respectively and the $M1$ transitions widths for $S$ and $P$ wave states are presented in Table \ref{tab:20}.

\section{\label{sec:S-D Mixing} S-D Mixing}

In bottomonium, the proximity of energy levels of higher excited states with same $J^{PC}$ can result in mixing of states. The mixing is caused by the tensor potential term, but it is not strong enough to induce substantial mixing \cite{58a,59}. However, for the states above open flavor threshold the mixing can be caused by coupled-channel dynamics, threshold effects, meson exchange, and multi-gluon exchange interactions \cite{59a,59b,59c,59d}. These effects can modifiy the wavefunctions, causing mass shift and mixing between states, as well as affecting decay properties such as open channel strong decay, leptonic decays etc. Consequently, the conventional representation of bottomonium states as pure $S$ and $D$ wavefunctions breaks down, and the states are instead identified as admixtures of both components. The mixed states can be represented in terms of pure $|nS\rangle$ and $|n'D\rangle$ states as \cite{59}
\begin{align}
	|\phi\rangle &= \cos\theta|nS\rangle + \sin\theta|n'D\rangle \,,
	\nonumber \\
	|\phi'\rangle &= -\sin\theta|nS\rangle + \cos\theta|n'D\rangle \,,
\end{align}
where $|\phi\rangle$ and $|\phi'\rangle$ are the mixed states, and $\theta$ is the mixing angle. The masses of the mixed states can be calculated using \cite{59}
\begin{align}
	M_{\phi} &= \left[\left(\frac{M_{nS}+M_{n'D}}{2}\right)+\left(\frac{M_{nS}-M_{n'D}}{2\cos2\theta}\right)\right] \,,
	\nonumber \\
	M_{\phi'} &= \left[\left(\frac{M_{nS}+M_{n'D}}{2}\right)+\left(\frac{M_{n'D}-M_{nS}}{2\cos2\theta}\right)\right] \,.
\end{align}
Here $M_{\phi}$ and $M_{\phi'}$ are the masses of the mixed states, and $M_{nS}$ and $M_{n'D}$ are the masses of the corresponding pure $S$ and $D$ states. 

\noindent The leptonic decay widths of the mixed states are given by \cite{59,60}
\begin{align}
	\Gamma_{\phi} &= \biggl[\frac{2\alpha e_{q}}{M_{nS}}|R_{nS}(0)|\cos\theta  +\frac{5\alpha e_{q}}{\sqrt{2}m_{q}^{2}M_{n'D}}|R''_{n'D}(0)|\sin\theta\biggr]^{2} \,,
	\nonumber \\
	\Gamma_{\phi'} &= \biggl[\frac{5\alpha e_{q}}{\sqrt{2}m_{q}^{2}M_{n'D}}|R''_{n'D}(0)|\cos\theta -\frac{2\alpha e_{q}}{M_{nS}}|R_{nS}(0)|\sin\theta\biggr]^{2}\,.
\end{align}
The leptonic decay of the mixed states is fitted to the experimental data to obtain the mixing angle, which is then used to calculate the masses of the mixed states.  Our results of $S-D$ mixing are presented in Table \ref{tab:21}.

\section{\label{sec:Results and Discussion} Results and Discussion}

In this study, a screened potential model within a relativistic framework is employed to compute the mass spectrum and decay widths of $b\bar{b}$ bound system. The masses of $S$-wave states are presented in Table \ref{tab:2} and are compared with the experimental data and other theoretical models. The well-established $1S$ and $2S$ states serve as benchmarks, with our model predicting $\eta_{b}(1S)= 9406.4$ MeV, $\Upsilon(1S)= 9451.1$ MeV, $\eta_{b}(2S)= 9998.9$ MeV and $\Upsilon(2S) = 10023.8$ MeV. The hyperfine mass splitting given by $\Delta m(nS) = m[\Upsilon(nS)] - m[\eta_b(nS)]$, are evaluated to be $\Delta m(1S) = 44.7$ MeV and $\Delta m(2S) = 24.9$ MeV. These values are consistent with the experimental results of $\Delta m(1S) = 62.3 \pm 3.2$ MeV and $\Delta m(2S) = 24.3 \pm 3.5_{-1.9}^{+2.8}$ MeV \cite{exp}. The $\Upsilon(10355)$ is well established as the $\Upsilon(3S)$ in the literature. In our model its mass is evaluated to be $10394.2$ MeV. Our model predicts the mass difference $m[\Upsilon(3S)]- m[\Upsilon(2S)]=370.4$ MeV compared to the experimental value of $331.50 \pm 0.02 \pm 0.13$ MeV \cite{exp}. The $\Upsilon(10580)$ is traditionally identified as the  $\Upsilon(4S)$ state \cite{55,57,58}. Our model calculates the mass of $\Upsilon(4S)$ as $10688.1$ MeV which is overestimated by $108.7$ MeV compared to experimental value. This overestimation is a consistent trend observed across all potential models \cite{57,58}. A $^{3}P_{0}$ model analysis suggests that $\Upsilon(10580)$ exhibits a significant meson-meson component due its proximity with the $B^{*}\bar{B}^{*}$ channel \cite{63}. Ref. \cite{64} suggests that the state discovered by the CLEO Collaboration at $10684 \pm 10 \pm 8$ MeV, identified as a $b\bar{b}g$ hybrid \cite{65}, is a more suitable assignment for the $\Upsilon(4S)$ state, which is also corroborated by our model. The intermediate $B^{*}\bar{B}^{*}$ channel may induce observable $S-D$ mixing within the $\Upsilon(10580)$ state \cite{63}, and Ref. \cite{59} predicts it to be $\Upsilon(4S)-\Upsilon(3D)$ mixture state with a substantial mixing angle. The $\Upsilon(10860)$ and $\Upsilon(11020)$ states are associated with the $\Upsilon(5S)$ and $\Upsilon(6S)$ states, respectively \cite{57}. Our model predicts their masses as $10938.9$ MeV and $11160.9$ MeV, which are overestimated by $53.7$ MeV and $160.9 MeV$, respectively. Theoretical models commonly show discrepancies in  $\Upsilon(5S)$ and $\Upsilon(6S)$ mass predictions, either overestimating or underestimating their values. Various interpretations have been explored in the literature, where $\Upsilon(10860)$ is considered as mixture of $\Upsilon(5S)- P$ wave hybrid \cite{66}, while lattice QCD studies remain inconclusive on whether $\Upsilon(11020)$ corresponds to $\Upsilon(S)$ or $\Upsilon(D)$ state \cite{67}. The $^{3}P_{0}$ model of Ref \cite{63} concluded that $\Upsilon(10860)$ and $\Upsilon(11020)$ are structures are mainly $b\bar{b}$ states with small $S-D$ mixing component. This was analyzed in Ref. \cite{68}, proposing $\Upsilon(10860)$ as a $\Upsilon(5S)-\Upsilon(4D)$ mixture, and in Ref. \cite{59} it is suggested that both $\Upsilon(10860)$ and $\Upsilon(11020)$ are $\Upsilon(5S)-\Upsilon(4D)$ mixture. More experimental data is required to understand their nature. We discuss the possibility of $S-D$ mixing in $\Upsilon(10580)$, $\Upsilon(10860)$ and $\Upsilon(11020)$ later in this section. 

\begin{table}
	\caption{\label{tab:2} $S$ wave mass spectra of $b\bar{b}$ states (in MeV)}
	\begin{ruledtabular}
	\begin{tabular}{cccccccc}
	States & Ours & Exp\cite{exp} & \cite{57} & \cite{52} & \cite{58} & \cite{55} & \cite{62}
	\\
	\hline
	$1^{1}S_{0}$ & 9406.4 & 9398.7 & 9398 & 9402 & 9423 & 9412.22 & 9390
	\\
	$2^{1}S_{0}$ & 9998.9 & 9999.0 & 9989 & 9976 & 9983 & 9995.48 & 9990
	\\
	$3^{1}S_{0}$ & 10374.9 &  & 10336 & 10336 & 10342 & 10339.00 & 10326
	\\
	$4^{1}S_{0}$ & 10671.8 &  & 10597 & 10635 & 10638 & 10572.49 & 10584
	\\
	$5^{1}S_{0}$ & 10924.5 &  & 10810 & 10869 & 10901 & 10746.76 & 10800
	\\
	$6^{1}S_{0}$ & 11147.9 &  & 10991 & 11097 & 11140 & 11064.47 & 10988
	\\
	\hline
	$1^{3}S_{1}$ & 9451.1 & 9460.3 & 9463 & 9465 & 9463 & 9460.75 & 9460
	\\
	$2^{3}S_{1}$ & 10023.8 & 10023.3 & 10017 & 10003 & 10001 & 10026.22 & 10015
	\\
	$3^{3}S_{1}$ & 10394.2 & 10355.1 & 10356 & 10354 & 10354 & 10364.65 & 10343
	\\
	$4^{3}S_{1}$ & 10688.1 & 10579.4 & 10612 & 10635 & 10650 & 10594.47 & 10597
	\\
	$5^{3}S_{1}$ & 10938.9 & 10885.2 & 10822 & 10878 & 10912 & 10766.14 & 10811
	\\
	$6^{3}S_{1}$ & 11160.9 & 11000.0 & 11001 & 11102 & 11151 & 11081.70 & 10997
	\end{tabular}
	\end{ruledtabular}
\end{table}

\begin{table}
	\caption{\label{tab:3} $P$ wave mass spectra of $b\bar{b}$ states (in MeV)}
	\begin{ruledtabular}
	\begin{tabular}{cccccccc}
	States & Ours & Exp\cite{exp} & \cite{57} & \cite{52} & \cite{58} & \cite{55} & \cite{62}
	\\
	\hline
	$1^{1}P_{1}$ & 9872.9 & 9899.3 & 9894 & 9882 & 9899 & 9874.56 & 9909 
	\\
	$2^{1}P_{1}$ & 10271.7 & 10259.8 & 10259 & 10250 & 10268 & 10270.00 & 10254 
	\\
	$3^{1}P_{1}$ & 10582.7 &  & 10530 & 10541 & 10570 & 10526.50 & 10519 
	\\
	$4^{1}P_{1}$ & 10845.6 &  & 10751 & 10790 &  & 10714.80 &  
	\\
	$5^{1}P_{1}$ & 11077.1 &  & 10938 & 11016 &  & 10863.00 &  
	\\
	\hline
	$1^{3}P_{0}$ & 9838.7 & 9859.4 & 9858 & 9847 & 9874 & 9849.61 & 9864 
	\\
	$2^{3}P_{0}$ & 10244.9 & 10232.5 & 10235 & 10226 & 10248 & 10252.54 & 10220 
	\\
	$3^{3}P_{0}$ & 10559.4 &  & 10513 & 10522 & 10551 & 10512.88 & 10490 
	\\
	$4^{3}P_{0}$ & 10824.5 &  & 10736 & 10775 &  & 10703.56 &  
	\\
	$5^{3}P_{0}$ & 11057.4 &  & 10926 & 11004 &  & 10853.38 &  
	\\
	\hline
	$1^{3}P_{1}$ & 9865.7 & 9892.8 & 9889 & 9876 & 9894 & 9871.47 & 9903 
	\\
	$2^{3}P_{1}$ & 10266.2 & 10255.5 & 10255 & 10246 & 10265 & 10267.86 & 10249 
	\\
	$3^{3}P_{1}$ & 10578.1 & 10513.4 & 10527 & 10538 & 10567 & 10524.84 & 10515 
	\\
	$4^{3}P_{1}$ & 10841.5 &  & 10749 & 10788 &  & 10713.44 &  
	\\
	$5^{3}P_{1}$ & 11073.3 &  & 10936 & 11014 &  & 10861.83 & 
	\\
	\hline
	$1^{3}P_{2}$ & 9885.6 & 9912.2 & 9910 & 9897 & 9907 & 9881.40 & 9921
	\\
	$2^{3}P_{2}$ & 10282.3 & 10268.6 & 10269 & 10261 & 10274 & 10274.77 & 10264
	\\
	$3^{3}P_{2}$ & 10592.3 & 10524.0 & 10539 & 10550 & 10576 & 10530.21 & 10528
	\\
	$4^{3}P_{2}$ & 10854.6 &  & 10758 & 10798 &  & 10717.86 & 
	\\
	$5^{3}P_{2}$ & 11085.6 &  & 10944 & 11022 &  & 10865.62 & 
	\end{tabular}
	\end{ruledtabular}
\end{table}

The $P$-wave masess are presented in Table \ref{tab:3} and our evaluated masses for $1P$ and $2P$ states are in accordance with the experimental values. The experimentally determined mass difference are $m[\chi_{b2}(1P)] - m[\chi_{b1}(1P)] = 19.10 \pm 0.25$ MeV, $m[\chi_{b1}(1P)] - m[\chi_{b0}(1P)] = 32.49 \pm 0.93$ MeV, $m[\chi_{b2}(2P)] - m[\chi_{b1}(2P)] = 13.10 \pm 0.24$ MeV and $m[\chi_{b1}(2P)] - m[\chi_{b0}(2P)] = 23.8 \pm 1.7$ MeV \cite{exp}. Our model calculates these values as $19.9$ MeV, $27$ MeV, $16.1$ MeV and $21.3$ MeV, respectively, exhibiting good agreement with the experimental data. Among $3P$ bottomonium states, only $\chi_{b1}(3P)$ and $\chi_{b2}(3P)$ have been identified. In our model their masses are obtained as $10578.1$ MeV and $10592.3$ MeV, which are higher by $64.7$ MeV and $68.9$ MeV, respectively. This discrepancy can be due to proximity to open-flavor $B\bar{B}^{*}$ threshold, potentially causing mixing effects \cite{69,70}. The experimentally measured mass difference $m[\chi_{b2}(3P)] - m[\chi_{b1}(3P)] = 10.60 \pm 0.64 \pm 0.17$ MeV \cite{exp} is calculated as $14.2$ MeV in our model. Masses of $D$-wave states are presented in Table \ref{tab:4}. The mass of $\Upsilon_{1}(1D)$ state in our model is evaluated to be $10147.9$ MeV, deviating by $15.8$ MeV from the experimental value \cite{exp}. The $\Upsilon_{2}(1D)$ and $\Upsilon_{3}(1D)$ states are estimated to have mass values of $10.13$ GeV and $10.18$ GeV, respectively \cite{15}, while our model calculates them as $10139.4$ MeV and $10154.2$ MeV, respectively. The recently observed $\Upsilon(10753)$ is generally associated with $\Upsilon_{1}(3D)$ \cite{57,70}, although alternative interpretations suggest a tetraquark \cite{71,72}, hybrid meson \cite{3},etc. The mass of $\Upsilon_{1}(3D)$ state in our model is evaluated to be $10741.3$ MeV, aligning with the experimental value \cite{exp}. A reanalysis of BABAR data estimated the mass of $\Upsilon_{1}(2D)$ to be $10495 \pm 5$ MeV  with $10.7\sigma$ significance \cite{73}, while our model calculates it as $10467.3$ MeV, showing consistency with experimental result. Masses of $F$ and $G$ - wave states are presented in Table \ref{tab:5}. Our model shows consistency with other models for lower states, but deviations arise for higher excitations. The masses for different $J$ states in Table \ref{tab:5} are very close to each other which could make it harder to differentiate these states experimentally.

\begin{table}
	\caption{\label{tab:4} $D$ wave mass spectra of $b\bar{b}$ states (in MeV)}
	\begin{ruledtabular}
	\begin{tabular}{cccccccc}
	States & Ours & Exp\cite{exp} & \cite{57} & \cite{52} & \cite{58} & \cite{55} & \cite{62}
	\\
	\hline 
	$1^{1}D_{2}$ & 10149.1 &  & 10163 & 10148 & 10149 & 10153.80 & 10153
	\\
	$2^{1}D_{2}$ & 10476.3 &  & 10450 & 10450 & 10465 & 10456.60 & 10432
	\\
	$3^{1}D_{2}$ & 10749.9 &  & 10681 & 10706 & 10740 & 10664.70 & 
	\\
	$4^{1}D_{2}$ & 10989.2 &  & 10876 & 10935 & 10988 & 10823.00 & 
	\\
	$5^{1}D_{2}$ & 11204.1 &  & 11046 &  &  & 10952.60 & 
	\\
	\hline
	$1^{3}D_{1}$ & 10139.4 &  & 10153 & 10138 & 10145 & 10144.99 & 10146
	\\
	$2^{3}D_{1}$ & 10467.3 &  & 10442 & 10441 & 10462 & 10450.23 & 10425
	\\
	$3^{3}D_{1}$ & 10741.3 & 10752.7 & 10675 & 10698 & 10736 & 10659.68 & 
	\\
	$4^{3}D_{1}$ & 10981.0 &  & 10871 & 10928 & 10985 & 10818.83 & 
	\\
	$5^{3}D_{1}$ & 11196.1 &  & 11041 &  &  & 10949.01 & 
	\\
	\hline
	$1^{3}D_{2}$ & 10147.9 & 10163.7 & 10162 & 10147 & 10149 & 10152.77 & 10153
	\\
	$2^{3}D_{2}$ & 10475.0 &  & 10450 & 10449 & 10465 & 10455.86 & 10432
	\\
	$3^{3}D_{2}$ & 10748.7 &  & 10681 & 10705 & 10740 & 10664.12 & 
	\\
	$4^{3}D_{2}$ & 10987.9 &  & 10876 & 10934 & 10988 & 10822.52 & 
	\\
	$5^{3}D_{2}$ & 11202.8 &  & 11045 &  &  & 10951.59 & 
	\\
	\hline
	$1^{3}D_{3}$ & 10154.2 &  & 10170 & 10155 & 10150 & 10158.31 & 10157
	\\
	$2^{3}D_{3}$ & 10481.1 &  & 10456 & 10455 & 10466 & 10459.85 & 10436
	\\
	$3^{3}D_{3}$ & 10754.6 &  & 10686 & 10711 & 10741 & 10667.25 & 
	\\
	$4^{3}D_{3}$ & 10993.7 &  & 10880 & 10939 & 10990 & 10825.12 & 
	\\
	$5^{3}D_{3}$ & 11208.5 &  & 11049 &  &  & 10954.42 & 
	\end{tabular}
	\end{ruledtabular}
\end{table}

\begin{table}
	\caption{\label{tab:5} $F$ and $G$ wave mass spectra of $b\bar{b}$ states (in MeV)}
	\begin{ruledtabular}
	\begin{tabular}{cccccccc}
	States & Ours & \cite{57} & \cite{52} & States & Ours & \cite{57} & \cite{52}
	\\
	\hline
	$1^{1}F_{3}$ & 10366.8 & 10366 & 10355 & $1^{1}G_{4}$ & 10552.9 & 10534 & 10530
	\\
	$2^{1}F_{3}$ & 10652.2 & 10609 & 10619 & $2^{1}G_{4}$ & 10809.8 & 10747 & 10770
	\\
	$3^{1}F_{3}$ & 10900.0 & 10812 & 10853 & $3^{1}G_{4}$ & 11038.2 & 10929 & 
	\\
	$4^{1}F_{3}$ & 11121.5 & 10988 &  & $4^{1}G_{4}$ & 11245.3 &  & 
	\\
	$5^{1}F_{3}$ & 11323.0 &  &  & $5^{1}G_{4}$ & 11435.6 &  & 
	\\
	\hline
	$1^{3}F_{2}$ & 10363.6 & 10362 & 10350 & $1^{3}G_{3}$ & 10552.8 & 10533 & 10529
	\\
	$2^{3}F_{2}$ & 10648.8 & 10605 & 10615 & $2^{3}G_{3}$ & 10809.2 & 10745 & 10769
	\\
	$3^{3}F_{2}$ & 10896.5 & 10809 & 10850 & $3^{3}G_{3}$ & 11037.3 & 10928 & 
	\\
	$4^{3}F_{2}$ & 11117.8 & 10985 &  & $4^{3}G_{3}$ & 11244.1 &  & 
	\\
	$5^{3}F_{2}$ & 11319.2 &  &  & $5^{3}G_{3}$ & 11434.3 &  & 
	\\
	\hline
	$1^{3}F_{3}$ & 10366.8 & 10366 & 10355 & $1^{3}G_{4}$ & 10553.4 & 10535 & 10531
	\\
	$2^{3}F_{3}$ & 10652.2 & 10609 & 10619 & $2^{3}G_{4}$ & 10810.1 & 10747 & 10770
	\\
	$3^{3}F_{3}$ & 10899.9 & 10812 & 10853 & $3^{3}G_{4}$ & 11038.4 & 10929 & 
	\\
	$4^{3}F_{3}$ & 11121.3 & 10988 &  & $4^{3}G_{4}$ & 11245.5 &  & 
	\\
	$5^{3}F_{3}$ & 11322.7 &  &  & $5^{3}G_{4}$ & 11435.8 &  & 
	\\
	\hline
	$1^{3}F_{4}$ & 10368.5 & 10369 & 10358 & $1^{3}G_{5}$ & 10552.6 & 10536 & 10532
	\\
	$2^{3}F_{4}$ & 10654.1 & 10612 & 10622 & $2^{3}G_{5}$ & 10809.8 & 10748 & 10772
	\\
	$3^{3}F_{4}$ & 10902.1 & 10815 & 10856 & $3^{3}G_{5}$ & 11038.5 & 10931 & 
	\\
	$4^{3}F_{4}$ & 11123.7 & 10990 &  & $4^{3}G_{5}$ & 11245.9 &  & 
	\\
	$5^{3}F_{4}$ & 11325.3 &  &  & $5^{3}G_{5}$ & 11436.4 &  & 
	\end{tabular}
	\end{ruledtabular}
\end{table}

\begin{table}
	\caption{\label{tab:6} Pseudoscalar and vector decay constants (in MeV)}
	\begin{ruledtabular}
	\begin{tabular}{ccccccc}
	States & $f_{P/V}$ & Exp\cite{exp} & \cite{58} & \cite{55} & \cite{74} & \cite{75}
	\\
	\hline
	$1^{1}S_{0}$ & 655.9 &  & 529 & 578.21 & 646.025 & 744 
	\\
	$2^{1}S_{0}$ & 489.2 &  & 317 & 499.48 & 518.803 & 577 
	\\
	$3^{1}S_{0}$ & 431.8 &  & 280 & 450.35 & 474.954 & 511 
	\\
	$4^{1}S_{0}$ & 398.6 &  & 264 & 413.93 & 449.654 & 471
	\\
	$5^{1}S_{0}$ & 375.5 &  & 255 & 385.68 & 432.072 & 443
	\\
	$6^{1}S_{0}$ & 357.6 &  & 249 & 360.93 & 418.645 & 422
	\\
	\hline
	$1^{3}S_{1}$ & 640.2 & 715$\pm$5 & 530 & 551.53 & 647.250 & 706 
	\\
	$2^{3}S_{1}$ & 478.0 & 498$\pm$8 & 317 & 477.05 & 519.436 & 547 
	\\
	$3^{3}S_{1}$ & 422.0 & 430$\pm$4 & 280 & 430.42 & 475.440 & 484 
	\\
	$4^{3}S_{1}$ & 389.7 & 336$\pm$18 & 265 & 395.80 & 450.066 & 446 
	\\
	$5^{3}S_{1}$ & 349.7 &  & 255 & 368.91 & 432.437 & 419 
	\\
	$6^{3}S_{1}$ & 335.3 &  & 249 & 345.40 & 418.977 & 399 
	\end{tabular}
	\end{ruledtabular}
\end{table}

\begin{table}
	\caption{\label{tab:7} Decay constants of $P$-wave states (in MeV)}
	\begin{ruledtabular}
	\begin{tabular}{cccc}
		States & $f_{\chi_{0}}$ & States & $f_{\chi_{1}}$ 
		\\
		\hline
		$1^{3}P_{0}$ & 227.8 & $1^{3}P_{1}$ & 262.4 
		\\
		$2^{3}P_{0}$ & 248.7 & $2^{3}P_{1}$ & 286.6  
		\\
		$3^{3}P_{0}$ & 257.3 & $3^{3}P_{1}$ & 296.6  
		\\
		$4^{3}P_{0}$ & 262.0 & $4^{3}P_{1}$ & 302.1  
		\\
		$5^{3}P_{0}$ & 264.8 & $5^{3}P_{1}$ & 305.4  
	\end{tabular}
	\end{ruledtabular}
\end{table}

\begin{table*}
	\caption{\label{tab:8} Di-leptonic decay widths (in keV for $S$ states and in eV for $D$ states)}
	\begin{ruledtabular}
	\begin{tabular}{ccccccccc}
	States & $\Gamma$ & $\Gamma_{cf}$ & Exp\cite{exp} & \cite{59} & \cite{57} & \cite{52} & \cite{55} & \cite{58}
	\\
	\hline
	$1^{3}S_{1}$ & 1.268 & 0.883 & 1.34$\pm$0.018 & 1.370 & 1.65 & 1.44 & 0.7700 & 0.582
	\\
	$2^{3}S_{1}$ & 0.666 & 0.464 & 0.612$\pm$0.011 & 0.626 & 0.821 & 0.73 & 0.5442 & 0.197
	\\
	$3^{3}S_{1}$ & 0.501 & 0.349 & 0.443$\pm$0.008 & 0.468 & 0.569 & 0.53 & 0.4288 & 0.149
	\\
	$4^{3}S_{1}$ & 0.415 & 0.289 & 0.272$\pm$0.029 & 0.393 & 0.431 & 0.39 & 0.3549 & 0.129
	\\
	$5^{3}S_{1}$ & 0.360 & 0.251 & 0.31$\pm$0.07 & 0.346 & 0.348 & 0.33 & 0.3035 & 0.117
	\\
	$6^{3}S_{1}$ & 0.320 & 0.223 & 0.13$\pm$0.03 & 0.313 & 0.286 & 0.27 & 0.2586 & 0.109
	\\
	\hline
	$1^{3}D_{1}$ & 1.149 &  &  & 2.0 & 1.88 & 1.38 & 5.0 & 1.65
	\\
	$2^{3}D_{1}$ & 2.166 &  &  & 3.0 & 2.81 & 1.99 & 5.8 & 2.42
	\\
	$3^{3}D_{1}$ & 3.059 &  &  & 5.0 & 3.00 & 2.38 & 5.9 & 3.19
	\\
	$4^{3}D_{1}$ & 4.573 &  &  & 6.0 & 3.00 & 2.18 & 5.8 & 3.97
	\\
	$5^{3}D_{1}$ & 5.219 &  &  & 8.0 & 3.02 &  & 5.7 &
	\end{tabular}
	\end{ruledtabular}
\end{table*}

Decay constants of pseudoscalar $(f_{P})$, vector $(f_{V})$ and tensor $(f_{\chi_{0}},f_{\chi_{1}})$ states are presented in Tables \ref{tab:6} and \ref{tab:7}, respectively. Our calculated values for the vector decay constants ($f_{V}$) are in accord with experimental values and shows more consistency over other theoretical models. The di-leptonic decay widths $\Gamma(l^{+}l^{-})$ of $\Upsilon(nS)$ and $\Upsilon(nD)$ states, without $(\Gamma)$ and with $(\Gamma_{cf})$ the correction factor are presented in Table \ref{tab:8}. The di-leptonic decay widths of $\Upsilon(nS)$ states evaluated without the correction term are more in agreement with the experimental value, while the correction factor significantly suppresses them. The di-leptonic decay widths of $\Upsilon(nD)$ are smaller than $\Upsilon(nS)$ by factor of $1000$, serving as a key distinguishing feature in most models \cite{55,57}. The di-leptonic decay width difference between $\Upsilon(nS)$ and $\Upsilon(nD)$ states is used as a justification for assigning $\Upsilon(10580)$, $\Upsilon(10860)$, and $\Upsilon(11020)$ states to the $\Upsilon(4S)$, $\Upsilon(5S)$, and $\Upsilon(6S)$, respectively, in potential models. Since  $\Upsilon(10580)$, $\Upsilon(10860)$ and $\Upsilon(11020)$ exhibit $S$ state characteristics rather than being purely $D$ state, their widths are often overestimated, suggesting a potential for $S-D$ mixing \cite{59}. For $n \ge 3$ the probability of $S-D$ mixing increases and even a small mixing angle can increase the di-leptonic decay widths of $\Upsilon(nD)$ by order of 2 \cite{76}.  To study $S-D$ mixing in our model, the di-leptonic decay widths without the correction factor are utilized to obtain the mixing angle. The di-photonic decay widths $\Gamma(\gamma \gamma)$ of bottomonium states without $(\Gamma)$ and with $(\Gamma_{cf})$ the correction factor are listed in Table \ref{tab:9}. Our results are comparable to Ref. \cite{55,74} in magnitude, but are lower than those in Ref. \cite{52,57}. The di-photonic decay width of $\eta_{b}(1S)$ is not seen experimentally and we predict it to be $0.344$ keV. The tri-photonic decay widths $\Gamma(\gamma \gamma \gamma)$ without $(\Gamma)$ and with $(\Gamma_{cf})$ the correction factor are listed in Table \ref{tab:10}. The values of tri-photonic decay widths vary significantly among models, highlighting the need for experimental validation. The di-gluonic decay widths $\Gamma(gg)$ without $(\Gamma)$ and with $(\Gamma_{cf})$ the correction factor are calculated in Table \ref{tab:11} and Table \ref{tab:12}. Our di-gluonic decay widths of $S$, $P$ and $D$ states are comparable to \cite{55} but are $2$-$4$ times smaller than those in \cite{52,57,58,77}. For lower-lying $\eta_{b}(nS)$ states, the di-gluonic decay widths constitute approximately $\sim 100\%$ of their total width due to suppression of OZI-allowed two-body strong decays \cite{57}. Our evaluated di-gluonic width for $\eta_{b}(1S)$ is $5.763$, close to the lower limit of total width estimate of $10.0_{-4}^{+5}$ MeV \cite{exp}. The tri-gluonic decay widths $\Gamma(ggg)$ without $(\Gamma)$ and with $(\Gamma_{cf})$ the correction factor are presented in Table \ref{tab:13}. The tri-gluonic decay width for $\Upsilon(1S)$ is lower by $13.95$ MeV, while those for $\Upsilon(2S)$ and $\Upsilon(3S)$ agree well with experimental results. For $P$ and $D$ states, our predicted widths are lower than other models. The photo-gluonic decay widths $\Gamma(\gamma gg)$ and quark-gluonic decay width $\Gamma(q\bar{q}g)$ without $(\Gamma)$ and with $(\Gamma_{cf})$ the correction factor are evaluated in Table \ref{tab:14}. The photo-gluonic decay widths of $\Upsilon(1S)$, $\Upsilon(2S)$ and $\Upsilon(3S)$ are in accordance with the experimental data. The multi-gluon or hybrid $q\bar{q}g$ decays are dominant channel for $\chi_{b1}(1P)$ state \cite{57}. Our predictions for quark-gluonic decay width for $\chi_{b1}(1P)$ are observed to be lower compared to other models.

\begin{table*}
	\caption{\label{tab:9} Di-photonic decay widths (in keV)}
	\begin{ruledtabular}
	\begin{tabular}{cccccccc}
	States & $\Gamma$ & $\Gamma_{cf}$ & \cite{57} & \cite{52} & \cite{55} & \cite{74} & \cite{58} 
	\\
	\hline
	$1^{1}S_{0}$ & 0.426 & 0.344 & 1.05 & 0.94 & 0.3035 & 0.387 & 0.2361 
	\\
	$2^{1}S_{0}$ & 0.223 & 0.180 & 0.489 & 0.41 & 0.2122 & 0.263 & 0.0896 
	\\
	$3^{1}S_{0}$ & 0.168 & 0.135 &  0.323 & 0.29 & 0.1668 & 0.229 & 0.0726
	\\
	$4^{1}S_{0}$ & 0.139 & 0.112 & 0.237 & 0.20 & 0.1378 & 0.212 & 0.0666
	\\
	$5^{1}S_{0}$ & 0.120 & 0.097 & 0.192 & 0.17 & 0.1176 & 0.201 & 0.0636
	\\
	$6^{1}S_{0}$ & 0.107 & 0.086 & 0.152 & 0.14 & 0.1000 & 0.193 & 0.0619
	\\
	\hline
	$1^{3}P_{0}$ & 0.042 & 0.042 & 0.199 & 0.15 & 0.1150 & 0.0196 & 0.0168 
	\\
	$2^{3}P_{0}$ & 0.046 & 0.047 & 0.205 & 0.15 & 0.1014 & 0.0195 & 0.0172 
	\\
	$3^{3}P_{0}$ & 0.046 & 0.047 & 0.180 & 0.13 & 0.0875 & 0.0194 & 0.0192 
	\\
	$4^{3}P_{0}$ & 0.045 & 0.046 & 0.157 & 0.13 & 0.0768 & 0.0192 &  
	\\
	$5^{3}P_{0}$ & 0.044 & 0.046 & 0.146 &  & 0.0686 & 0.0191 &  
	\\
	\hline
	$1^{3}P_{2}$ & 0.011 & 0.007 & 0.0106 & 0.0093 & 0.0147 & 0.0052 & 0.0024 
	\\
	$2^{3}P_{2}$ & 0.012 & 0.008 & 0.0133 & 0.012 & 0.0131 & 0.0052 & 0.0025
	\\
	$3^{3}P_{2}$ & 0.012 & 0.009 &  0.0141 & 0.013 & 0.0114 & 0.0051 & 0.0027 
	\\
	$4^{3}P_{2}$ & 0.012 & 0.008 & 0.0142 & 0.015 & 0.0100 & 0.0051 & 
	\\
	$5^{3}P_{2}$ & 0.011 & 0.008 & 0.0143 &  & 0.0090 & 0.0050 & 
	\end{tabular}
	\end{ruledtabular}
\end{table*}

\begin{table*}
	\caption{\label{tab:10} Tri-photonic decay widths (in $10^{-3}$ eV)}
	\begin{ruledtabular}
	\begin{tabular}{ccccccc}
	States & $\Gamma$ & $\Gamma_{cf}$ & \cite{57} & \cite{52} & \cite{77} & \cite{58} 
	\\
	\hline
	$1^{3}S_{1}$ & 42.16 & 11.92 & 19.4 & 17.0 & 3.44 & 30.67 
	\\
	$2^{3}S_{1}$ & 22.17 & 6.27 & 10.9 & 9.8 & 2.00 & 11.58 
	\\
	$3^{3}S_{1}$ & 16.66 & 4.71 & 8.04 & 7.6 & 1.55 & 9.376 
	\\
	$4^{3}S_{1}$ & 13.81 & 3.91 & 6.36 & 6.0 & 1.29 & 8.590 
	\\
	$5^{3}S_{1}$ & 11.98 & 3.39 & 5.43 &  & 1.10 & 8.206 
	\\
	$6^{3}S_{1}$ & 10.65 & 3.01 & 4.57 &  & 0.96 & 7.982 
	\end{tabular}
	\end{ruledtabular}
\end{table*}

\begin{table*}
	\caption{\label{tab:11} Di-gluonic decay widths of $S$, $P$ (in MeV) and $D$ (in keV) states}
	\begin{ruledtabular}
	\begin{tabular}{cccccccc}
	States & $\Gamma$ & $\Gamma_{cf}$ & \cite{57} & \cite{52} & \cite{77} & \cite{58} & \cite{55} 
	\\
	\hline
	$1^{1}S_{0}$ & 4.608 & 5.763 & 17.9 & 16.6 & 20.18 & 11.326 & 6.8520 
	\\
	$2^{1}S_{0}$ & 2.412 & 3.016 & 8.33 & 7.2 & 10.64 & 4.301 & 5.2374 
	\\
	$3^{1}S_{0}$ & 1.811 & 2.264 & 5.51 & 4.9 & 7.94 & 3.485 & 4.3182 
	\\
	$4^{1}S_{0}$ & 1.500 & 1.876 & 4.03 & 3.4 &  & 3.193 & 3.6829
	\\
	$5^{1}S_{0}$ & 1.301 & 1.627 & 3.26 &  &  & 3.051 & 3.2196
	\\
	$6^{1}S_{0}$ & 1.156 & 1.446 & 2.59 &  &  & 2.968 & 2.8519
	\\
	\hline
	$1^{3}P_{0}$ & 0.454 & 0.713 & 3.37 & 2.6 & 2.00 & 1.34 & 1.4297 
	\\
	$2^{3}P_{0}$ & 0.499 & 0.783 & 3.52 & 2.6 & 2.37 & 1.39 & 1.2358 
	\\
	$3^{3}P_{0}$ & 0.503 & 0.789 & 3.10 & 2.2 & 2.46 & 1.54 & 1.0539 
	\\
	$4^{3}P_{0}$ & 0.496 & 0.779 & 2.73 & 2.1 &  &  &  0.9175
	\\
	$5^{3}P_{0}$ & 0.486 & 0.763 & 2.54 &  &  &  & 0.8127 
	\\
	\hline
	$1^{3}P_{2}$ & 0.119 & 0.118 & 0.165 & 0.147 & 0.837 & 0.209 & 0.2370 
	\\
	$2^{3}P_{2}$ & 0.131 & 0.130 & 0.220 & 0.207 & 0.104 & 0.215 & 0.2064 
	\\
	$3^{3}P_{2}$ & 0.132 & 0.132 & 0.243 & 0.227 & 0.111 & 0.240 & 0.1767 
	\\
	$4^{3}P_{2}$ & 0.131 & 0.130 & 0.251 & 0.248 &  &  & 0.1543
	\\
	$5^{3}P_{2}$ & 0.128 & 0.127 & 0.258 &  &  &  & 0.1370
	\\
	\hline
	$1^{1}D_{2}$ & 0.321 & 0.281 & 0.657 & 1.8 & 0.37 & 0.489 & 
	\\
	$2^{1}D_{2}$ & 0.534 & 0.468 & 1.22 & 3.3 & 0.67 & 0.764 & 
	\\
	$3^{1}D_{2}$ & 0.679 & 0.595 & 1.59 & 4.7 &  & 1.06 & 
	\\
	$4^{1}D_{2}$ & 0.785 & 0.686 & 1.86 &  &  & 1.38 & 
	\\
	$5^{1}D_{2}$ & 0.861 & 0.754 & 2.13 &  &  &  & 
	\end{tabular}
	\end{ruledtabular}
\end{table*} 

\begin{table}
	\caption{\label{tab:12} Di-gluonic decay widths of $F$ (in keV) and $G$ (in eV) states}
	\begin{ruledtabular}
	\begin{tabular}{cccccccc}
	States & $\Gamma$ & \cite{57} & \cite{52} & States & $\Gamma$ & \cite{57} & \cite{52} 
	\\
	\hline
	$1^{3}F_{2}$ & 0.282 & 0.834 & 0.70 & $1^{3}F_{4}$ & 0.031 & 0.05 & 0.048
	\\
	$2^{3}F_{2}$ & 0.618 & 2.04 & 1.77 & $2^{3}F_{4}$ & 0.067 & 0.126 & 0.13
	\\
	$3^{3}F_{2}$ & 0.946 & 3.17 &  & $3^{3}F_{4}$ & 0.102 & 0.210 &
	\\
	$4^{3}F_{2}$ & 1.248 &  &  & $4^{3}F_{4}$ & 0.135 &  &
	\\
	$5^{3}F_{2}$ & 1.517 &  &  & $5^{3}F_{4}$ & 0.164 &  &
	\\
	\hline
	$1^{3}F_{3}$ & 0.031 &  0.0672 & 0.060 & $1^{1}G_{4}$ & 0.289 & 0.661 & 2.3
	\\
	$2^{3}F_{3}$ & 0.067 & 0.167 & 0.16 &
	$2^{1}G_{4}$ & 0.778 &  &
	\\
	$3^{3}F_{3}$ & 0.103 & 0.270 &  &
	$3^{1}G_{4}$ & 1.383 &  &
	\\
	$4^{3}F_{3}$ & 0.135 &  &  &
	$4^{1}G_{4}$ & 2.044 &  &
	\\
	$5^{3}F_{3}$ & 0.165 &  &  &
	$5^{1}G_{4}$ & 2.723 &  &
	\end{tabular}
	\end{ruledtabular}
\end{table}

\begin{table}
	\caption{\label{tab:13}Tri-gluonic decay widths (in keV)}
	\begin{ruledtabular}
	\begin{tabular}{cccccccc}
	States & $\Gamma$ & $\Gamma_{cf}$ & Exp\cite{exp} & \cite{57} & \cite{52} & \cite{55} & \cite{77} 
	\\
	\hline
	$1^{3}S_{1}$ & 41.85 & 30.18 & 44.13$\pm$1.09 & 50.8 & 47.6 & 28.5 & 41.63
	\\
	$2^{3}S_{1}$ & 22.00 & 15.86 & 18.8$\pm$1.59 & 28.4 & 26.3 & 19.3 & 24.25 
	\\
	$3^{3}S_{1}$ & 16.54 & 11.92 & 7.25$\pm$0.85 & 21.0 & 19.8 & 14.8 & 18.76
	\\
	$4^{3}S_{1}$ & 13.71 & 9.89 &  & 16.7 & 15.1 & 12.1 & 15.58 
	\\
	$5^{3}S_{1}$ & 11.89 & 8.58 &  & 14.2 & 13.1 & 10.2 & 13.33 
	\\
	$6^{3}S_{1}$ & 10.57 & 7.62 &  & 12.0 & 11.0 & 8.5 & 11.57 
	\\
	\hline
	$1^{1}P_{1}$ & 20.10 &  &  & 44.7 & 37.0 & 35.7 & 35.26 
	\\
	$2^{1}P_{1}$ & 26.99 &  &  & 64.6 & 54.0 & 34.6 & 52.70 
	\\
	$3^{1}P_{1}$ & 30.23 &  &  & 71.1 & 59.0 & 33.1 & 62.16 
	\\
	$4^{1}P_{1}$ & 31.99 &  &  & 73.2 & 64.0 & 32.7 & 
	\\
	$5^{1}P_{1}$ & 32.97 &  &  & 76.2 &  & 30.9 & 
	\\
	\hline
	$1^{3}D_{1}$ & 3.11 &  &  & 10.4 & 8.11 & 10.6 & 9.97 
	\\
	$2^{3}D_{1}$ & 5.61 &  &  & 20.1 & 14.8 & 11.9 & 9.69 
	\\
	$3^{3}D_{1}$ & 7.54 &  &  & 26.0 & 21.2 & 11.8 & 
	\\
	$4^{3}D_{1}$ & 9.05 &  &  & 30.4 &  & 11.3 & 
	\\
	$5^{3}D_{1}$ & 1.02 &  &  & 34.7 &  & 10.8 & 
	\\
	\hline
	$1^{3}D_{2}$ & 0.37 &  &  & 0.821 & 0.69 &  & 0.62
	\\
	$2^{3}D_{2}$ & 0.66 &  &  & 1.65 & 1.4 &  & 0.61
	\\
	$3^{3}D_{2}$ & 0.89 &  &  & 2.27 & 2.0 &  &
	\\
	$4^{3}D_{2}$ & 0.11 &  &  & 2.75 &  &  &
	\\
	$5^{3}D_{2}$ & 0.12 &  &  & 3.23 &  &  &
	\\
	\hline
	$1^{3}D_{3}$ & 1.46 &  &  & 2.19 & 2.07 & 6.0 & 0.22 
	\\
	$2^{3}D_{3}$ & 2.64 &  &  & 4.56 & 4.3 & 5.6 & 1.25 
	\\
	$3^{3}D_{3}$ & 3.54 &  &  & 6.65 & 6.6 & 5.5 & 
	\\
	$4^{3}D_{3}$ & 4.26 &  &  & 8.38 &  & 5.3 & 
	\\
	$5^{3}D_{3}$ & 4.82 &  &  & 10.1 &  & 5.1 & 
	\end{tabular}
	\end{ruledtabular}
\end{table}

\begin{table}
	\caption{\label{tab:14}Photo-gluon decay widths of $S$ states and quark-gluon decay widths of $P$ states (in keV)}
	\begin{ruledtabular}
	\begin{tabular}{cccccccc}
	States & $\Gamma$ & $\Gamma_{cf}$ & Exp\cite{exp} & \cite{57} & \cite{77} & \cite{58} & \cite{55} 
	\\
	\hline
	$1^{3}S_{1}$ & 1.37 & 0.79 & 1.19$\pm$0.33 & 1.32 & 0.79 & 0.903 & 0.7220
	\\
	$2^{3}S_{1}$ & 0.72 & 0.42 & 0.60$\pm$0.10 & 0.739 & 0.46 & 0.341 & 0.4982
	\\
	$3^{3}S_{1}$ & 0.54 & 0.31 & 0.20$\pm$0.04 & 0.547 & 0.36 & 0.276 & 0.3874
	\\
	$4^{3}S_{1}$ & 0.45 & 0.26 &  & 0.433 & 0.30 & 0.253 & 0.3176
	\\
	$5^{3}S_{1}$ & 0.39 & 0.22 &  & 0.370 & 0.25 & 0.242 & 0.2698
	\\
	$6^{3}S_{1}$ & 0.34 & 0.19 &  & 0.311 & 0.22 & 0.235 & 0.2272
	\\
	\hline
	$1^{3}P_{1}$ & 32.25 &  &  & 81.7 & 71.53 & 45.55 & 57.9585
	\\
	$2^{3}P_{1}$ & 43.28 &  &  & 117.0 & 106.14 & 56.16 & 55.3966
	\\
	$3^{3}P_{1}$ & 48.46 &  &  & 126.0 & 124.53 & 68.97 & 52.9585
	\\
	$4^{3}P_{1}$ & 51.27 &  &  & 128.0 &  &  & 52.4466
	\\
	$5^{3}P_{1}$ & 52.82 &  &  & 132.0 &  &  & 49.5181
	\end{tabular}
	\end{ruledtabular}
\end{table}

The $S$ wave $E1$ transitions widths are calculated in Table \ref{tab:15}. The transition widths $\Gamma(2S \rightarrow \gamma \chi_{b}(1P))$ in our model align well with experimental data. The transition widths for $\Gamma(\Upsilon(3S) \rightarrow \gamma \chi_{b}(P))$ presents a complex scenario, due to discrepancies in $\Gamma(\Upsilon(3S) \rightarrow \gamma \chi_{b}(2P))$ and $\Gamma(\Upsilon(3S) \rightarrow \gamma \chi_{b}(1P))$ predictions across models \cite{57,77}. Our model estimates $\Gamma(\Upsilon(3S) \rightarrow \gamma \chi_{b}(2P))$ slightly higher than experimental values, while $\Gamma(\Upsilon(3S) \rightarrow \gamma \chi_{b}(1P))$ are highly suppressed, a typical feature of $E1$ transitions among states separated by two radial nodes, making them susceptible to relativistic corrections \cite{78,79}. This suppression is evident in our evaluation, where $\Gamma(\Upsilon(3S) \rightarrow \gamma \chi_{b0}(1P)) = 0.057$ keV and $\Gamma(\Upsilon(3S) \rightarrow \gamma \chi_{b2}(1P)) = 0.139$ align with experimental results, while $\Gamma(\Upsilon(3S) \rightarrow \gamma \chi_{b1}(1P)) = 0.115$ keV exceeds the experimental value. This atypical hierarchy of $\Gamma(\Upsilon(3S) \rightarrow \gamma \chi_{b2}(1P))>\Gamma(\Upsilon(3S) \rightarrow \gamma \chi_{b0}(1P))>\Gamma(\Upsilon(3S) \rightarrow \gamma \chi_{b1}(1P))$ mentioned in Ref. \cite{80} is also observed in our model. This is attributed to $\chi_{b1}(1P)$ mixing with $\chi_{b}(2P)$ and $\chi_{b}(3P)$, further suppressing $\Gamma(\Upsilon(3S) \rightarrow \gamma \chi_{b1}(1P))$. In Ref. \cite{59}, the $S-D$ mixing in $\Upsilon(3S)$ is considered to explain the $E1$ transitions widths of $\Gamma(\Upsilon(3S) \rightarrow \gamma \chi_{b}(2P))$, which allowed them to reproduce the experimental widths. This explanation may also be extended for analysis of $\Gamma(\Upsilon(3S) \rightarrow \gamma \chi_{b}(1P))$ transition widths. We also evaluate the $\Gamma(4S \rightarrow \gamma P)$. The $P$ wave $E1$ transitions widths are presented in Table \ref{tab:16} and \ref{tab:17}. The transition width $\Gamma(1P \rightarrow \gamma S)$ in our model agrees with the experimental and theoretical results. Using the measured branching ratios $B[\chi_{b0}(1P) \rightarrow \gamma \Upsilon(1S)]=1.94 \pm 0.27 \%$, $B[\chi_{b1}(1P) \rightarrow \gamma \Upsilon(1S)]=35.2 \pm 2.0 \%$, and $B[\chi_{b2}(1P) \rightarrow \gamma \Upsilon(1S)]=18.0 \pm 1.0 \%$ \cite{exp}, we calculate the total decay width as $1.19$ MeV for $\chi_{b0}(1P)$, $79.0$ keV for $\chi_{b1}(1P)$, and $177.0$ keV for $\chi_{b2}(1P)$. Our total width for $\chi_{b0}(1P)$ is consistent with $1.3 \pm 0.9$ MeV and $\Gamma_{total}<2.4$ MeV condition predicted by Belle Collaboration \cite{81}. The $h_{b}(1P)$ has primary transition $h_{b}(1P) \rightarrow \gamma \eta_{b}(1S)$ with measured branching ratio of $52_{-5}^{+6} \%$. Using this, we estimate the total decay width of $h_{b}(1P)$ as $74.0$ keV, consistent with the Ref. \cite{62}. The evaluated transition width $h_{b}(2P) \rightarrow \gamma \eta_{b}(2S)$ is lower than the experimental value, a trend seen in most of the potential models. Using measured branching ratios $B[h_{b}(2P) \rightarrow \gamma \eta_{b}(2S)]=48 \pm 13 \%$ and $B[h_{b}(2P) \rightarrow \gamma \eta_{b}(1S)]=22 \pm 5 \%$ \cite{exp}, we estimate the total decay width of $h_{b}(2P)$ as $46.0$ keV and $50.0$ keV, respectively, with an average of $48.0$ keV, which is smaller than the estimate in Ref. \cite{62}. The transition width $\Gamma(\chi_{b0}(1P) \rightarrow \gamma \Upsilon(2S))$ is overestimated in most models. From the measured branching ratios $B[\chi_{b0}(2P) \rightarrow \gamma \Upsilon(2S)]= 1.38 \pm 0.30 \%$ and $B[\chi_{b0}(2P) \rightarrow \gamma \Upsilon(1S)]=(3.8 \pm 1.7) \times 10^{-3}$ \cite{exp}, we determine the total decay width of $\chi_{b0}(2P)$ as $0.88$ MeV and $2.20$ MeV, respectively. While these values vary significantly, the latter aligns with the ($\sim 2.5$ MeV) prediction of Ref. \cite{52}. Our model predicts $\Gamma(\chi_{b1}(2P) \rightarrow \gamma \Upsilon(S))$ and $\Gamma(\chi_{b2}(2P) \rightarrow \gamma \Upsilon(S))$ in excellent agreement with experimental results. Using measured branching ratios $B[\chi_{b1}(2P) \rightarrow \gamma \Upsilon(2S)]= 18.1 \pm 1.9 \%$ and $B[\chi_{b1}(2P) \rightarrow \gamma \Upsilon(1S)]= 9.9 \pm 1.0 \%$ \cite{exp}, we determine the total decay width of $\chi_{b1}(2P)$ as $87.0$ keV and $91.0$ keV, respectively,with an average of $89.0$ keV, which aligns with the CLEO Collaboration value $96 \pm 16$ keV \cite{82}. Using the branching ratio $B[\chi_{b2}(2P) \rightarrow \gamma \Upsilon(1S)]= 6.6 \pm 0.8 \%$ \cite{exp}, we calculate the decay width of $146.0 \pm 18.0$ keV for $\chi_{b2}(2P)$, which agrees with $138 \pm 19$ keV obtained by CLEO Collaboration \cite{82}. No experimental data exists for $\Gamma(3P \rightarrow \gamma S)$ and $\Gamma(3P \rightarrow \gamma D)$ transitions, although detections of  $\Gamma(\chi_{b1}(3P) \rightarrow \gamma \Upsilon(1S))$, $\Gamma(\chi_{b1}(3P) \rightarrow \gamma \Upsilon(2S))$, $\Gamma(\chi_{b1}(3P) \rightarrow \gamma \Upsilon(3S))$, and $\Gamma(\chi_{b2}(3P) \rightarrow \gamma \Upsilon(3S))$ have been reported. We estimate these transition widths as $4.754$ keV, $5.876$ keV, $12.895$ keV, and $15.899$ keV, respectively. $E1$ transitions widths for $D$-wave states are presented in Table \ref{tab:18}. The transition $\Gamma(\Upsilon_{2}(1D) \rightarrow \gamma \chi_{b}(1P))$ has been observed \cite{exp} and our model estimates $\Gamma(\Upsilon_{2}(1D) \rightarrow \gamma \chi_{b2}(1P)) = 6.194$ keV and $\Gamma(\Upsilon_{2}(1D) \rightarrow \gamma \chi_{b1}(1P))= 22.890$ keV are consistent with other theoretical models. Ref. \cite{52,57,77} suggest the total decay widths of $\eta_{b}(1D)$ and $\Upsilon_{3}(1D)$ are equivalent to their transition widths $\Gamma(\eta_{b}(1D) \rightarrow \gamma h_{b}(1P))$ and $\Gamma(\Upsilon_{3}(1D) \rightarrow \gamma h_{b}(1P))$, respectively. We estimate these transition widths to be $28.719$ keV and $26.518$ keV, respectively, in accordance with other models. The $\Upsilon_{1}(1D)$ state is predicted to be detected in $\Gamma(\Upsilon_{1}(1D) \rightarrow \gamma \chi_{b0}(1P))$ and $\Gamma(\Upsilon_{1}(1D) \rightarrow \gamma \chi_{b1}(1P))$ due to their large branching ratios \cite{52,57,77}. Our model estimates these transition widths as $20.292$ keV and $11.661$ keV, respectively. The large branching ratio of $\Gamma(\eta_{b}(2D) \rightarrow \gamma h_{b}(2P))$ suggests that the unobserved $\eta_{b}(2D)$ state could be detected \cite{52,57}. Our model estimates $\Gamma(\eta_{b}(2D) \rightarrow \gamma h_{b}(2P))$ to be $21.128$ keV. The transition widths of $2D$ states in our model align with other theoretical predictions. $E1$ transitions widths for $F$ and $G$ wave states are listed in Table \ref{tab:19}, which are slightly higher than those in other potential models. The $M1$ transition widths are presented in Table \ref{tab:20}. Our $M1$ transition widths show noticeable differences from Refs. \cite{52,55,57}. While our $\Gamma(\Upsilon(nS) \rightarrow \gamma \eta_{b}(nS))$ estimates are lower than other models, they aligns more closely with experimental results. Since decay widths  are highly dependent on the wavefunction, estimates vary significantly across models.

\begin{table}
	\caption{\label{tab:15}$E1$ transition widths (in keV) and photon energies (in MeV) of $S$ wave states}
	\begin{ruledtabular}
	\begin{tabular}{ccccccccc}
	Initial & Final & Ours & Ours  & Exp\cite{exp} & \cite{57} & \cite{52} & \cite{77} & \cite{62} 
	\\
	State & State & $E_{\gamma}$ & $\Gamma_{E1}$ &  &  &  &  & 
	\\
	\hline
	$2^{1}S_{0}$ & $1^{1}P_{1}$ & 125.3 & 4.769 &  & 2.467 & 2.48 & 2.85 & 3.41
	\\
	\hline
	$2^{3}S_{1}$ & $1^{3}P_{0}$ & 183.4 & 1.632 & 1.22$\pm$0.11 & 0.907 & 0.91 & 1.09 & 1.09
	\\
	& $1^{3}P_{1}$ & 156.9 & 3.092 & 2.21$\pm$0.22 & 1.60 & 1.63 & 1.84 & 2.17
	\\
	& $1^{3}P_{2}$ & 137.7 & 3.472 & 2.29$\pm$0.22 & 1.86 & 1.88 & 2.08 & 2.62
	\\
	\hline
	$3^{1}S_{0}$ & $2^{1}P_{1}$ & 102.7 & 6.596 &  & 2.88 & 2.96 & 2.60 & 4.25
	\\
	& $1^{1}P_{1}$ & 489.9 & 0.226 &  & 1.12 & 1.3 & 0.0084 & 0.67
	\\
	\hline
	$3^{3}S_{1}$ & $2^{3}P_{0}$ & 148.2 & 2.157 & 1.20$\pm$0.12 & 1.06 & 1.03 & 1.21 & 1.21
	\\
	& $2^{3}P_{1}$ & 127.2 & 4.132 & 2.56$\pm$0.26 & 1.96 & 1.91 & 2.13 & 2.61
	\\
	& $2^{3}P_{2}$ & 111.3 & 4.651 & 2.66$\pm$0.27 & 2.37 & 2.30 & 2.56 & 3.16
	\\
	& $1^{3}P_{0}$ & 540.6 & 0.057 & 0.055$\pm$0.01 & 0.0099 & 0.01 & 0.15 & 0.097
	\\
	& $1^{3}P_{1}$ & 515.1 & 0.115 & 0.018$\pm$0.01 & 0.0363 & 0.05 & 0.16 & 0.0005
	\\
	& $1^{3}P_{2}$ & 496.1 & 0.139 & 0.2$\pm$0.03 & 0.359 & 0.45 & 0.0827 & 0.14
	\\
	\hline
	$4^{1}S_{0}$ & $3^{1}P_{1}$ & 88.6 & 7.329 &  & 1.50 & 1.24 &  & 
	\\
	& $2^{1}P_{1}$ & 392.5 & 0.718 &  &  & 0.732 &  & 
	\\
	& $1^{1}P_{1}$ & 768.9 & 0.022 &  & 0.688 &  &  & 
	\\
	\hline
	$4^{3}S_{1}$ & $3^{3}P_{0}$ & 127.9 & 2.384 &  & 0.587 & 0.48 & 0.61 & 
	\\
	& $3^{3}P_{1}$ & 109.5 & 4.544 &  & 1.14 & 0.84 & 1.17 & 
	\\
	& $3^{3}P_{2}$ & 95.4 & 5.057 &  & 1.16 & 0.82 & 1.45 & 
	\\
	& $2^{3}P_{0}$ & 434.0 & 0.160 &  & 0.0137 &  & 0.17 & 
	\\
	& $2^{3}P_{1}$ & 413.6 & 0.344 &  & 0.0138 &  & 0.18 & 
	\\
	& $2^{3}P_{2}$ & 398.1 & 0.440 &  & 0.226 &  & 0.11 & 
	\\
	& $1^{3}P_{0}$ & 815.7 & 0.007 &  & 5.12$\times10^{-4}$ &  & 0.0588 & 
	\\
	& $1^{3}P_{1}$ & 790.8 & 0.012 &  & 0.0507 &  & 0.0474 & 
	\\
	& $1^{3}P_{2}$ & 772.4 & 0.013 &  & 0.219 &  & 0.012 & 
	\end{tabular}
	\end{ruledtabular}
\end{table}

\begin{table*}
	\caption{\label{tab:16}$E1$ transition widths (in keV) and photon energies (in MeV) of $1P$ and $2P$ wave states}
	\begin{ruledtabular}
	\begin{tabular}{ccccccccc}
	Initial & Final & Ours & Ours  & Exp\cite{exp} & \cite{57} & \cite{52} & \cite{77} & \cite{62} 
	\\
	State & State & $E_{\gamma}$ & $\Gamma_{E1}$ &  &  &  &  & 
	\\
	\hline
	$1^{1}P_{1}$ & $1^{1}S_{0}$ & 455.4 & 38.692 & 35.77 & 34.4 & 35.7 & 43.66 & 35.8 
	\\
	\hline
	$1^{3}P_{0}$ & $1^{3}S_{1}$ & 379.9 & 23.099 &  & 22.8 & 23.8 & 28.07 & 27.5 
	\\
	\hline
	$1^{3}P_{1}$ & $1^{3}S_{1}$ & 405.8 & 27.901 & 32.544 & 28.3 & 29.5 & 35.66 & 31.9 
	\\
	\hline
	$1^{3}P_{2}$ & $1^{3}S_{1}$ & 424.9 & 31.805 & 34.38 & 31.4 & 32.8 & 39.15 & 31.8 
	\\
	\hline
	$2^{1}P_{1}$ & $1^{1}D_{2}$ & 121.8 & 3.604 &  & 1.81 & 1.7 & 5.36 & 2.24 
	\\
	& $2^{1}S_{0}$ & 269.1 & 21.962 & 40.32 & 15.0 & 14.1 & 17.60 & 16.2 
	\\
	& $1^{1}S_{0}$ & 828.8 & 11.071 &  & 10.8 & 13.0 & 14.90 & 16.1 
	\\
	\hline
	$2^{3}P_{0}$ & $1^{3}D_{1}$ & 104.9 & 2.316 &  & 1.05 & 1.0 & 0.74 & 1.77 
	\\
	& $2^{3}S_{1}$ & 218.7 & 12.165 & 1.2$\times10^{-4}$ & 11.1 & 10.9 & 12.80 & 14.4 
	\\
	& $1^{3}S_{1}$ & 763.0 & 8.198 &  & 2.31 & 2.5 & 5.44 & 5.54 
	\\
	\hline
	$2^{3}P_{1}$ & $1^{3}D_{1}$ & 126.0 & 0.995 &  & 0.511 & 0.5 & 0.41 & 0.56 
	\\
	& $1^{3}D_{2}$ & 117.6 & 2.436 &  & 1.25 & 1.2 & 1.26 & 0.50 
	\\
	& $2^{3}S_{1}$ & 239.6 & 15.790 & 19.4$\pm$5 & 13.7 & 13.3 & 15.89 & 15.3 
	\\
	& $1^{3}S_{1}$ & 782.7 & 8.991 & 8.9$\pm$2.2 & 5.09 & 5.5 & 9.13 & 10.8
	\\
	\hline
	$2^{3}P_{2}$ & $1^{3}D_{1}$ & 141.9 & 0.056 &  & 0.0267 & 0.03 & 0.0209 & 0.026 
	\\
	& $1^{3}D_{2}$ & 133.5 & 0.708 &  & 0.339 & 0.3 & 0.35 & 0.42 
	\\
	& $1^{3}D_{3}$ & 127.3 & 3.442 &  & 1.61 & 1.5 & 2.06 & 2.51 
	\\
	& $2^{3}S_{1}$ & 255.2 & 18.908 & 15.1$\pm$5.6 & 14.6 & 14.3 & 17.50 & 15.3 
	\\
	& $1^{3}S_{1}$ & 797.6 & 9.626 & 9.8$\pm$2.3 & 7.86 & 8.4 & 11.38 & 12.5 
	\end{tabular}
	\end{ruledtabular}
\end{table*}

\begin{table*}
	\caption{\label{tab:17}$E1$ transition widths (in keV) and photon energies (in MeV) of $3P$ states}
	\begin{ruledtabular}
	\begin{tabular}{cccccccc}
	Initial & Final & Ours & Ours  & \cite{57} & \cite{52} & \cite{77} & \cite{62} 
	\\
	State & State & $E_{\gamma}$ & $\Gamma_{E1}$ &  &  &  &  
	\\
	\hline
	$3^{1}P_{1}$ & $2^{1}D_{2}$ & 105.9 & 5.482 & 1.44 & 1.6 & 4.72 & 4.21 
	\\
	& $1^{1}D_{2}$ & 424.7 & 0.208 & 0.0585 & 0.081 & 0.35 & 0.17 
	\\
	& $3^{1}S_{0}$ & 205.8 & 18.156 & 9.94 & 8.9 & 12.27 & 14.1 
	\\
	& $2^{1}S_{0}$ & 567.7 & 7.175 & 4.60 & 8.2 & 6.86 & 7.63 
	\\
	& $1^{1}S_{0}$ & 1110.9 & 5.592 & 3.91 & 3.6 & 7.96 & 10.7 
	\\
	\hline
	$3^{3}P_{0}$ & $2^{3}D_{1}$ & 91.8 & 3.593 & 0.966 & 1.0 & 3.50 & 2.20 
	\\
	& $1^{3}D_{1}$ & 411.7 & 0.163 & 0.189 & 0.20 & 3.59$\times10^{-2}$ & 0.15 
	\\
	& $3^{3}S_{1}$ & 163.9 & 9.527 & 7.15 & 6.9 & 8.50 & 7.95 
	\\
	& $2^{3}S_{1}$ & 522.0 & 5.156 & 1.26 & 1.7 & 2.99 & 2.55 
	\\
	& $1^{3}S_{1}$ & 1050.1 & 4.462 & 0.427 & 0.3 & 1.99 & 1.87 
	\\
	\hline
	$3^{3}P_{1}$ & $2^{3}D_{1}$ & 110.2 & 1.541 & 0.425 & 0.47 & 1.26 & 1.07 
	\\
	& $2^{3}D_{2}$ & 102.5 & 3.738 & 0.950 & 1.1 & 3.34 & 0.94 
	\\
	& $1^{3}D_{1}$ & 429.6 & 0.056 & 0.00418 & 7.0$\times10^{-3}$ & 4.80$\times10^{-2}$ & 0.010 
	\\
	& $1^{3}D_{2}$ & 421.4 & 0.147 & 0.0615 & 0.080 & 0.11 & 0.015 
	\\
	& $3^{3}S_{1}$ & 182.3 & 12.897 & 8.36 & 8.4 & 9.62 & 10.3 
	\\
	& $2^{3}S_{1}$ & 539.8 & 5.876 & 2.49 & 3.1 & 4.58 & 5.63 
	\\
	& $1^{3}S_{1}$ & 1066.9 & 4.754 & 1.62 & 1.3 & 4.17 & 6.41 
	\\
	\hline
	$3^{3}P_{2}$ & $2^{3}D_{1}$ & 124.3 & 0.088 & 0.0248 & 0.027 & 0.18 & 0.049
	\\
	& $2^{3}D_{2}$ & 116.6 & 1.090 & 0.295 & 0.32 & 0.79 & 0.78
	\\
	& $2^{3}D_{3}$ & 110.6 & 5.229 & 1.37 & 1.5 & 4.16 & 4.60
	\\
	& $1^{3}D_{1}$ & 443.2 & 0.003 & 1.15$\times10^{-4}$ &  & 3.38$\times10^{-3}$ & 0.047
	\\
	& $1^{3}D_{2}$ & 435.1 & 0.037 & 3.11$\times10^{-4}$ &  & 4.41$\times10^{-2}$ & 0.068
	\\
	& $1^{3}D_{3}$ & 429.0 & 0.188 & 0.0288 & 0.046 & 0.21 & 0.12
	\\
	& $3^{3}S_{1}$ & 196.2 & 15.889 & 9.30 & 9.3 & 10.38 & 10.8
	\\
	& $2^{3}S_{1}$ & 553.2 & 6.478 & 3.66 & 4.5 & 5.62 & 6.72
	\\
	& $1^{3}S_{1}$ & 1079.7 & 4.986 & 3.17 & 2.8 & 5.65 & 8.17
	\end{tabular}
	\end{ruledtabular}
\end{table*}

\begin{table*}
	\caption{\label{tab:18}$E1$ transition widths (in keV) and photon energies (in MeV) of $D$ states}
	\begin{ruledtabular}
	\begin{tabular}{cccccccc}
	Initial & Final & Ours & Ours  & \cite{57} & \cite{52} & \cite{77} & \cite{62} 
	\\
	State & State & $E_{\gamma}$ & $\Gamma_{E1}$ &  &  &  & 
	\\
	\hline
	$1^{1}D_{2}$ & $1^{1}P_{1}$ & 272.5 & 28.719 & 24.3 & 24.9 & 17.23 & 30.3  
	\\
	\hline
	$1^{3}D_{1}$ & $1^{3}P_{0}$ & 296.3 & 20.292 & 16.3 & 16.5 & 20.98 & 19.8 
	\\
	& $1^{3}P_{1}$ & 270.0 & 11.661 & 9.51 & 9.7 & 12.29 & 13.3 
	\\
	& $1^{3}P_{2}$ & 250.6 & 0.626 & 0.550 & 0.56 & 0.65 & 1.02 
	\\
	\hline
	$1^{3}D_{2}$ & $1^{3}P_{1}$ & 278.3 & 22.890 & 18.8 & 19.2 & 21.95 & 21.8 
	\\
	& $1^{3}P_{2}$ & 258.9 & 6.194 & 5.49 & 5.6 & 6.23 & 7.23 
	\\
	\hline
	$1^{3}D_{3}$ & $1^{3}P_{2}$ & 265.1 & 26.518 & 23.9 & 24.3 & 24.74 & 32.1 
	\\
	\hline
	$2^{1}D_{2}$ & $1^{1}F_{3}$ & 108.9 & 2.640 & 1.35 & 1.8 & 2.20 &  
	\\
	& $2^{1}P_{1}$ & 202.6 & 21.128 & 16.8 & 16.5 & 11.66 & 15.6 
	\\
	& $1^{1}P_{1}$ & 586.0 & 6.313 & 3.36 & 3.0 & 4.15 & 5.66 
	\\
	\hline
	$2^{3}D_{1}$ & $1^{3}F_{2}$ & 103.1 & 2.246 & 1.18 & 1.6 & 2.05 &   
	\\
	& $2^{3}P_{0}$ & 220.0 & 14.846 & 11.0 & 10.6 & 8.35 & 9.58 
	\\
	& $2^{3}P_{1}$ & 199.1 & 8.380 & 6.71 & 6.5 & 4.84 & 6.74 
	\\
	& $2^{3}P_{2}$ & 183.3 & 0.441 & 0.40 & 0.4 & 0.24 & 0.47 
	\\
	& $1^{3}P_{0}$ & 609.7 & 4.118 & 2.99 & 2.9 & 3.52 & 5.56
	\\
	& $1^{3}P_{1}$ & 584.3 & 2.599 & 1.03 & 0.9 & 1.58 & 2.17
	\\
	& $1^{3}P_{2}$ & 565.5 & 0.152 & 0.030 & 0.02 & 0.061 & 0.44
	\\
	\hline 
	$2^{3}D_{2}$ & $1^{3}F_{2}$ & 110.8 & 0.308 & 0.164 & 0.21 & 0.24 & 
	\\
	& $1^{3}F_{3}$ & 107.6 & 2.265 & 1.21 & 1.5 & 1.93 &  
	\\
	& $2^{3}P_{1}$ & 206.7 & 16.793 & 13.1 & 12.7 & 9.10 & 11.4 
	\\
	& $2^{3}P_{2}$ & 190.9 & 4.462 & 3.96 & 3.8 & 2.55 & 3.75 
	\\
	& $1^{3}P_{1}$ & 591.6 & 4.921 & 2.81 & 2.6 & 3.43 & 4.00 
	\\
	& $1^{3}P_{2}$ & 572.9 & 1.441 & 0.489 & 0.4 & 0.80 & 1.11 
	\\
	\hline
	$2^{3}D_{3}$ & $1^{3}F_{2}$ & 116.8 & 0.007 & 0.004 & 0.005 & 0.005 &  
	\\
	& $1^{3}F_{3}$ & 113.6 & 0.237 & 0.125 & 0.16 & 0.19 &  
	\\
	& $1^{3}F_{4}$ & 112.0 & 2.632 & 1.37 & 1.7 &  &  
	\\
	& $2^{3}P_{2}$ & 196.9 & 19.488 & 16.8 & 16.4 & 10.70 & 17.0 
	\\
	& $1^{3}P_{2}$ & 578.6 & 5.997 & 2.99 & 2.6 & 3.80 & 5.22 
	\end{tabular}
	\end{ruledtabular}
\end{table*}

\begin{table*}
	\caption{\label{tab:19}$E1$ transition widths (in keV) and photon energies (in MeV) of $F$ and $G$ states.}
	\begin{ruledtabular}
	\begin{tabular}{cccccc}
	Initial & Final & Ours & Ours & \cite{57} & \cite{52}   
	\\
	State & State & $E_{\gamma}$ & $\Gamma_{E1}$ &  &  
	\\
	\hline
	$1^{1}F_{3}$ & $1^{1}D_{2}$ & 215.3 & 27.505 & 22.0 & 18.8 
	\\
	\hline
	$1^{3}F_{2}$ & $1^{3}D_{1}$ & 221.8 & 25.149 & 19.4 & 16.4 
	\\
	& $1^{3}D_{2}$ & 213.4 & 4.171 & 3.26 & 2.7 
	\\
	& $1^{3}D_{3}$ & 207.3 & 0.109 & 0.0852 & 0.070 
	\\
	\hline
	$1^{3}F_{3}$ & $1^{3}D_{2}$ & 216.6 & 24.873 & 19.7 & 16.7 
	\\
	& $1^{3}D_{3}$ & 210.4 & 28.592 & 2.26 & 1.9 
	\\
	\hline
	$1^{3}F_{4}$ & $1^{3}D_{3}$ & 212.0 & 26.299 & 21.2 & 18.0 
	\\
	\hline
	$2^{1}F_{3}$ & $1^{1}G_{4}$ & 98.8 & 2.069 & 1.06 & 1.5 
	\\
	& $2^{1}D_{2}$ & 174.5 & 22.171 & 17.4 & 19.9 
	\\
	& $1^{1}D_{2}$ & 491.2 & 4.759 & 1.99 & 1.6 
	\\
	\hline
	$2^{3}F_{2}$ & $1^{3}G_{3}$ & 95.6 & 1.877 & 0.946 & 1.4 
	\\
	& $2^{3}D_{1}$ & 180.0 & 20.371 & 15.1 & 17.5 
	\\
	& $2^{3}D_{2}$ & 172.4 & 3.332 & 2.55 & 3.0 
	\\
	& $2^{3}D_{3}$ & 166.4 & 0.086 & 0.0681 & 0.080 
	\\
	& $1^{3}D_{1}$ & 497.2 & 4.214 & 1.95 & 1.6 
	\\
	& $1^{3}D_{2}$ & 489.1 & 0.727 & 0.224 & 0.16 
	\\
	& $1^{3}D_{3}$ & 483.1 & 0.019 & 0.00367 & 0.002 
	\\
	\hline
	$2^{3}F_{3}$ & $1^{3}G_{3}$ & 98.9 & 0.129 & 0.0664 & 0.10 
	\\
	& $1^{3}G_{4}$ & 98.2 & 1.910 & 0.957 & 1.4 
	\\
	& $2^{3}D_{2}$ & 175.6 & 20.091 & 15.4 & 17.9 
	\\
	& $2^{3}D_{3}$ & 169.7 & 2.275 & 1.80 & 2.1 
	\\
	& $1^{3}D_{2}$ & 492.3 & 4.273 & 1.83 & 1.4 
	\\
	& $1^{3}D_{3}$ & 486.3 & 0.506 & 0.145 & 0.1
	\\
	\hline
	$2^{3}F_{4}$ & $1^{3}G_{3}$ & 100.8 & 0.002 & 8.80$\times10^{-4}$ & 0.001 
	\\
	& $1^{3}G_{4}$ & 100.2 & 0.104 & 0.0535 & 0.080 
	\\
	& $1^{3}G_{5}$ & 100.9 & 2.098 & 1.05 & 1.5 
	\\
	& $2^{3}D_{3}$ & 171.6 & 21.148 & 16.9 & 19.6 
	\\
	& $1^{3}D_{3}$ & 488.2 & 4.635 & 0.126 & 1.4 
	\\
	\hline
	$1^{1}G_{4}$ & $1^{1}F_{3}$ & 184.5 & 27.129 & 21.1 & 23.1 
	\\
	\hline
	$1^{3}G_{3}$ & $1^{3}F_{2}$ & 187.5 & 26.098 & 20.1 & 22.3 
	\\
	& $1^{3}F_{3}$ & 184.3 & 2.174 & 1.67 & 1.8 
	\\
	& $1^{3}F_{4}$ & 182.7 & 0.034 & 0.0256 & 0.028 
	\\
	\hline
	$1^{3}G_{4}$ & $1^{3}F_{3}$ & 184.9 & 25.586 & 20.1 & 22.0 
	\\
	& $1^{3}F_{4}$ & 183.3 & 0.034 & 0.0256 & 0.028 
	\\
	\hline
	$1^{3}G_{5}$ & $1^{3}F_{4}$ & 182.6 & 26.304 & 21.1 & 23.1 
	\end{tabular}
	\end{ruledtabular}
\end{table*}

\begin{table*}
	\caption{\label{tab:20}$M1$ transition widths (in keV) and photon energies (in MeV) of $S$ and $P$ states}
	\begin{ruledtabular}
	\begin{tabular}{cccccccc}
	Initial & Final & Ours & Ours  & Exp\cite{exp} & \cite{57} & \cite{52} & \cite{77}  
	\\
	State & State & $E_{\gamma}$ & $\Gamma_{M1}$ &  &  &  &  
	\\
	\hline
	$1^{3}S_{1}$ & $1^{1}S_{0}$ & 44.5 & 4.228 &  & 9.52 & 10.0 & 9.34
	\\
	\hline
	$2^{1}S_{0}$ & $1^{3}S_{1}$ & 532.8 & 4.848 &  & 70.6 & 68.0 & 45.0
	\\
	\hline
	$2^{3}S_{1}$ & $2^{1}S_{0}$ & 24.8 & 0.732 &  & 0.582 & 0.590 & 0.580
	\\
	& $1^{1}S_{0}$ & 598.3 & 3.569 & 12.5$\pm$4.9 & 68.8 & 81.0 & 56.50
	\\
	\hline
	$3^{1}S_{0}$ & $2^{3}S_{1}$ & 345.1 & 2.015 &  & 11.1 & 9.10 & 9.20
	\\
	& $1^{3}S_{1}$ & 882.6 & 5.342 &  & 73.2 & 74.0 & 5.10
	\\
	\hline
	$3^{3}S_{1}$ & $3^{1}S_{0}$ & 19.3 & 0.343 &  & 0.337 & 0.250 & 0.658
	\\
	& $2^{1}S_{0}$ & 387.7 & 1.488 & $<$13 & 11.8 & 19.0 & 11.0
	\\
	& $1^{1}S_{0}$ & 940.8 & 2.836 & 10$\pm$2 & 60.4 & 60.0 & 57.0
	\\
	\hline
	$2^{1}P_{1}$ & $1^{3}P_{0}$ & 423.9 & 0.439 &  & 5.56 & 0.320 & 36.40
	\\
	& $1^{3}P_{1}$ & 398.0 & 0.857 &  & 1.30 & 1.10 & 1.280
	\\
	& $1^{3}P_{2}$ & 378.9 & 1.018 &  & 0.992 & 2.20 & 0.007
	\\
	\hline
	$2^{3}P_{0}$ & $1^{1}P_{1}$ & 365.2 & 0.475 &  & 5.21 & 9.70 & 2.390
	\\
	\hline
	$2^{3}P_{1}$ & $1^{1}P_{1}$ & 385.8 & 0.692 &  & 3.90$\times10^{-6}$ & 2.20 & 0.167
	\\
	\hline
	$2^{3}P_{2}$ & $1^{1}P_{1}$ & 401.2 & 0.905 &  & 3.86 & 0.240 & 1.780
	\\
	\hline
	$3^{1}P_{1}$ & $2^{3}P_{0}$ & 332.5 & 0.366 &  & 2.16 &  & 1.710
	\\
	& $2^{3}P_{1}$ & 311.8 & 0.709 &  & 0.559 &  & 0.597
	\\
	& $2^{3}P_{2}$ & 269.2 & 0.831 &  & 0.407 &  & 0.007
	\\
	& $1^{3}P_{0}$ & 717.9 & 0.279 &  & 5.10 & 0.980 & 3.770
	\\
	& $1^{3}P_{1}$ & 692.8 & 0.639 &  & 1.01 & 0.930 & 1.230
	\\
	& $1^{3}P_{2}$ & 674.2 & 0.867 &  & 1.48 & 0.140 & 0.051
	\\
	\hline
	$3^{3}P_{0}$ & $2^{1}P_{1}$ & 283.4 & 0.372 &  & 2.05 &  & 
	\\
	& $1^{1}P_{1}$ & 664.2 & 0.465 &  & 6.23 &  & 
	\\
	\hline
	$3^{3}P_{1}$ & $2^{1}P_{1}$ & 301.9 & 0.569 &  & 9.80$\times10^{-4}$ &  & 
	\\
	& $1^{1}P_{1}$ & 681.7 & 0.566 &  & 0.032 &  & 
	\\
	\hline 
	$3^{3}P_{2}$ & $2^{1}P_{1}$ & 315.7 & 0.772 &  & 1.74 &  & 
	\\
	& $1^{1}P_{1}$ & 694.9 & 0.654 &  & 3.53 &  & 
	\end{tabular}
	\end{ruledtabular}
\end{table*}

Table \ref{tab:21} presents the masses and leptonic decay widths of $S-D$ mixed states, which are assigned to experimentally observed states. The $\Upsilon(10355)$ state is considered as $3S-2D$ mixed state with a small mixing component, having mass of $10374.9$ MeV and leptonic width of $0.440$ keV. The $\Upsilon(10580)$ and $\Upsilon(10753)$ are considered to be $4S-3D$ mixed state with substantial mixing component, with mass of $10656.4$ MeV and leptonic width of $0.272$ keV for $\Upsilon(10580)$ and mass of $10772.9$ MeV and leptonic width of $0.129$ keV for $\Upsilon(10753)$. The $\Upsilon(10860)$ and $\Upsilon(11020)$ are assigned to be $5S-4D$ mixed states, which is also supported by Ref. \cite{59}. The mass and leptonic decay width values of the $\Upsilon(10860)$ and $\Upsilon(11020)$ are consistent with the experimental results. After taking into account the $S-D$ mixing, our final assignments are presented in Table \ref{tab:22}.

\begin{table*}
	\caption{\label{tab:21}$S-D$ mixed states with the masses of mixed states $M_{\phi}$ and $M_{\phi'}$ (in MeV) and their di-leptonic decay widths $\Gamma_{\phi}$ and $\Gamma_{\phi'}$ (in keV)}
	\begin{ruledtabular}
		\begin{tabular}{cccccccc}
		$S-D$ & $M_{S}$ & $\theta$ & $\theta$ \cite{59} & $M_{\phi}$ & $M_{exp}$ & $\Gamma_{\phi}$ & $\Gamma_{exp}$ 
		\\
		States & $M_{D}$ &  &  & $M_{\phi'}$ & \cite{exp} & $\Gamma_{\phi'}$ & \cite{exp}
		\\
		\hline
		$3S$ & 10394.2 & 19.28 & -9.0 & 10374.9 & 10355.1 & 0.440 & 0.443$\pm$0.008 \\
		$2D$ & 10467.3 &  &  & 10486.5 &  & 0.036 &  
		\\
		\hline
		$4S$ & 10688.1 & -28.82 & -12.5 &  10656.4 & 10579.4 & 0.272 & 0.272$\pm$0.029 
		\\
		$3D$ & 10741.3 &  &  & 10772.9 & 10752.7 & 0.129 &  
		\\
		\hline
		$5S$ & 10938.9 & 44.55 & -38.0 &  10909.8 & 10885.2 & 0.291 & 0.31$\pm$0.07 \\
		$4D$ & 10981.0 &  &  & 11010.1 & 11000.0 & 0.142 & 0.13$\pm$0.03 
		\end{tabular}
	\end{ruledtabular}
\end{table*}

\begin{table*}
	\caption{\label{tab:22}Our assignments for $b\bar{b}$ states with masses (in MeV) and di-leptonic decay widths (in keV)}
	\begin{ruledtabular}
	\begin{tabular}{cccccc}
	States & Assignment & $M_{exp}$ \cite{exp} & $M_{cal}$ & $\Gamma_{exp}^{ee}$ \cite{exp} & $\Gamma_{cal}^{ee}$ \\
	\hline
	$\eta_{b}(1S)$ & $\eta_{b}(1S)$ & 9398.7$\pm$2 & 9406.4 &  &  
	\\
	$\Upsilon(1S)$ & $\Upsilon(1S)$ & 9460.4$\pm$0.09$\pm$0.04 & 9451.1 & 1.34$\pm$0.018 & 1.268 
	\\
	$\chi_{b0}(1P)$ & $\chi_{b0}(1P)$ & 9859.44$\pm$0.42$\pm$0.31 & 9838.7 &  &  
	\\
	$\chi_{b1}(1P)$ & $\chi_{b1}(1P)$ & 9892.78$\pm$0.26$\pm$0.31 & 9865.7 &  &  
	\\
	$h_{b}(1P)$ & $h_{b}(1P)$ & 9899.3$\pm$0.8 & 9872.9 &  &  
	\\
	$\chi_{b2}(1P)$ & $\chi_{b2}(1P)$ & 9912.21$\pm$0.26$\pm$0.31 & 9885.6 &  &  
	\\
	$\eta_{b}(2S)$ & $\eta_{b}(2S)$ & 9999.0$\pm3.5_{-1.9}^{+2.8}$ & 9998.9 &  &  
	\\
	$\Upsilon(2S)$ & $\Upsilon(2S)$ & 10023.4$\pm$0.5 & 10023.3 & 0.612$\pm$0.011 & 0.666 
	\\
	$\Upsilon_{2}(1D)$ & $\Upsilon_{2}(1D)$ & 10163.7$\pm$1.4 & 10147.9 &  &  
	\\
	$\chi_{b0}(2P)$ & $\chi_{b0}(2P)$ & 10232.5$\pm$0.4$\pm$0.5 & 10244.9 &  &  
	\\
	$\chi_{b1}(2P)$ & $\chi_{b1}(2P)$ & 10255.46$\pm$0.22$\pm$0.5 & 10266.2 &  &  
	\\
	$h_{b}(2P)$ & $h_{b}(2P)$ & 10259.8$\pm$0.5$\pm$1.1 & 10271.7 &  &  
	\\
	$\chi_{b2}(2P)$ & $\chi_{b2}(2P)$ & 10268.65$\pm$0.22$\pm$0.5 & 10282.3 &  &  
	\\
	$\Upsilon(10355)$ & $\Upsilon(3S)-\Upsilon(2D)$ & 10355.1$\pm$0.5 & 10374.9 & 0.443$\pm$0.008 & 0.440 
	\\
	$\chi_{b1}(3P)$ & $\chi_{b1}(3P)$ & 10513.42$\pm$0.41$\pm$0.53 & 10578.1 &  & 
	\\
	$\chi_{b2}(3P)$ & $\chi_{b2}(3P)$ & 10524.02$\pm$0.57$\pm$0.53 & 10592.3 &  & 
	\\
	$\Upsilon(10580)$ & $\Upsilon(4S)-\Upsilon(3D)$ & 10579.4$\pm$1.2 & 10656.4 & 0.272$\pm$0.029 & 0.272 
	\\
	$\Upsilon(10753)$ & $\Upsilon(4S)-\Upsilon(3D)$ & 10752.7$\pm5.9_{-1.1}^{+0.7}$ & 10772.9 &  & 0.129  
	\\
	$\Upsilon(10860)$ & $\Upsilon(5S)-\Upsilon(4D)$ & $10885.2_{-1.6}^{+2.6}$ & 10909.8 & 0.31$\pm$0.07 & 0.291  
	\\
	$\Upsilon(11020)$ & $\Upsilon(5S)-\Upsilon(4D)$ & 11000$\pm$4 & 11010.1 & 0.13$\pm$0.03 & 0.142 
	\end{tabular}
	\end{ruledtabular}
\end{table*}

\clearpage

\section{\label{sec:Conclusion} Conclusion}

In this study, we explored the bottomonium system using a screened potential model within a relativistic framework to compute the mass spectrum of $S,P,D,F$ and $G$ wave, decay widths, and $E1$ and $M1$ transition widths, along with mass and leptonic decay widths of $S-D$ mixed states. This study emphasizes the relevance of relativistic dynamics, screening, and state mixing, offering a framework that bridges gaps between theory and experiment. The computed mass values exhibit strong agreement with experimental data, particularly for lower states, while our predictions for higher excited states demonstrate notable improvements compared to previous potential models. A recurring challenge in bottomonium spectroscopy has been reconciling theoretical predictions with experimental measurements, particularly for the masses and leptonic decay widths of higher states such as $\Upsilon(10355)$, $\Upsilon(10580)$,$\Upsilon(10860)$ and $\Upsilon(11020)$. Our study addresses this by incorporating $S-D$  mixing, yielding results that align closely with experimental values and providing a more refined interpretation of these states and emphasizing the need for beyond-static-potential effects in quarkonium spectroscopy. Our calculations of decay constants, particularly for vector states, show improvements over prior models along with annihilation decay widths. Beyond mass spectra, our evaluation of $E1$ and $M1$ transition widths offers valuable insights into radiative decays, supporting experimental searches for unobserved bottomonium states. We also utilized $E1$ transition widths to estimate the total decay widths of higher bottomonium states, achieving reasonable agreement with experimental values and reinforcing the validity of our model. Additionally, the calculated transition widths for higher excited states serve as references. This research paves the way for future investigations, particularly in the exploration of higher excited states and the effects of $S-D$ mixing.

\nocite{*}

\bibliography{BottomoniumCPCReference}

\end{document}